\documentclass[journal=aamick,manuscript=article]{achemso}%
\pdfoutput=1%
\usepackage[T1]{fontenc} % Use modern font encodings
\usepackage{xspace}%
\usepackage[percent]{overpic}%
\newcommand*{\CPB}{CsPbBr\textsubscript{3}\xspace}%
\newcommand*{\wtpct}{\,wt.\,\%\xspace}%
\newcommand*{\srcpath}{src}%
\newcommand*{\ts}[1]{\textsubscript{#1}}%
\newcommand*{\tS}[1]{\textsuperscript{#1}}%
\author{Tassilo Naujoks}%
\affiliation[Augsburg University]{Institut f\"{u}r Physik, Universität Augsburg, 86135 Augsburg, Germany}%
\author{Roshini Jayabalan}%
\affiliation[Augsburg University]{Institut f\"{u}r Physik, Universität Augsburg, 86135 Augsburg, Germany}%
\author{Christopher Kirsch}%
\affiliation[Tuebingen University]{Institut f\"{u}r Physikalische und Theoretische Chemie, Universität T\"{u}bingen, 72076 T\"{u}bingen, Germany}%
\author{Fengshuo Zu}%
\affiliation[Berlin University]{Institut f\"{u}r Physik \& IRIS Adlershof, Humboldt-Universit\"{a}t zu Berlin, 12489 Berlin, Germany}%
\author{Mukunda Mandal}%
\affiliation[Max Plank Institute Mainz]{Max Planck Institute f\"{u}r Polymerforschung, Ackermannweg 10, 55128 Mainz, Germany}%
\author{Jan Wahl}%
\affiliation[Tuebingen University]{Institut f\"{u}r Physikalische und Theoretische Chemie, Universität T\"{u}bingen, 72076 T\"{u}bingen, Germany}%
\author{Martin Waibel}%
\affiliation[Augsburg University]{Institut f\"{u}r Physik, Universität Augsburg, 86135 Augsburg, Germany}%
\author{Andreas Opitz}%
\affiliation[Berlin University]{Institut f\"{u}r Physik \& IRIS Adlershof, Humboldt-Universit\"{a}t zu Berlin, 12489 Berlin, Germany}%
\author{Norbert Koch}%
\affiliation[Berlin University]{Institut f\"{u}r Physik \& IRIS Adlershof, Humboldt-Universit\"{a}t zu Berlin, 12489 Berlin, Germany}%
\alsoaffiliation[Berlin Helmholtz]{Helmholtz-Zentrum Berlin für Materialien und Energie GmbH, 12489 Berlin, Germany}%
\author{Denis Andrienko}%
\affiliation[Max Plank Institute Mainz]{Max Planck Institute f\"{u}r Polymerforschung, Ackermannweg 10, 55128 Mainz, Germany}%
\author{Marcus Scheele}%
\affiliation[Tuebingen University]{Institut f\"{u}r Physikalische und Theoretische Chemie, Universität T\"{u}bingen, 72076 T\"{u}bingen, Germany}%
\author{Wolfgang Br\"utting}%
\email{wolfgang.bruetting@physik.uni-augsburg.de}%
\affiliation[Augsburg University]{Institut f\"{u}r Physik, Universität Augsburg, 86135 Augsburg, Germany}%
\date{\today}%
\title[LiTFSI-enhanced LHP NC LEDs]{Quantum Efficiency Enhancement of Lead-Halide Perovskite Nanocrystal LEDs by Organic Lithium Salt Treatment}%
\keywords{Perovskite nanocrystals, CsPbBr3 nanocrystals, LiTFSI doping, Perovskite LEDs}%
\begin{document}%
	\begin{tocentry}%
		\includegraphics[width=0.775\textwidth]{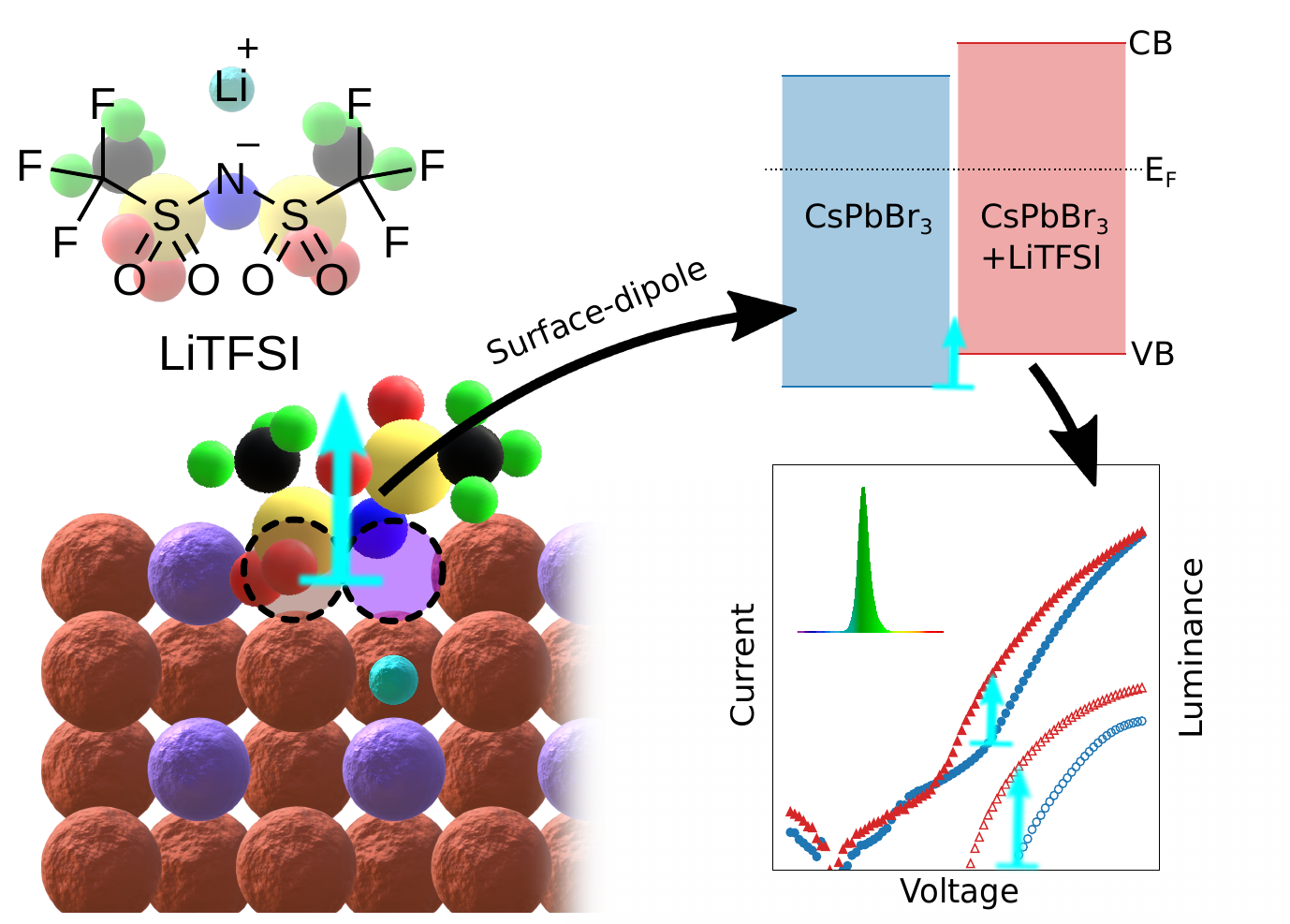}%
	\end{tocentry}%
	\begin{abstract}%
		Surface-defect passivation is key to achieving high photoluminescence quantum yield in lead halide perovskite nanocrystals. %
		However, in perovskite light-emitting diodes these surface ligands also have to enable balanced charge injection into the nanocrystals to yield high efficiency and operational lifetime. %
		In this respect, alkaline halides have been reported to passivate surface trap states and increase the overall stability of perovskite light emitters. %
		On the one side, the incorporation of alkaline ions into the lead halide perovskite crystal structure is considered to counterbalance cation vacancies, while on the other side, the excess halides are believed to stabilise the colloids. %
		Here, we report an organic lithium salt, viz. LiTFSI, as a halide-free surface passivation %
		on perovskite nanocrystals. %
		We show that treatment with LiTFSI has multiple beneficial effects on lead halide perovskite nanocrystals and LEDs derived from them. %
		We obtain higher photoluminescence quantum yield and longer exciton lifetime, and a radiation pattern that is more favourable for light outcoupling. %
		The ligand-induced dipoles on the nanocrystal surface shift their energy levels toward lower hole-injection barrier. %
		Overall, these effects add up to a four- to seven-fold boost of the external quantum efficiency in proof-of-concept LED structures, depending on the colour of the used lead halide perovskite nanocrystal emitters.%
	\end{abstract}%
	\maketitle%
	\section{Introduction}%
\begin{figure}[p]%
		\includegraphics[width=0.7\linewidth]{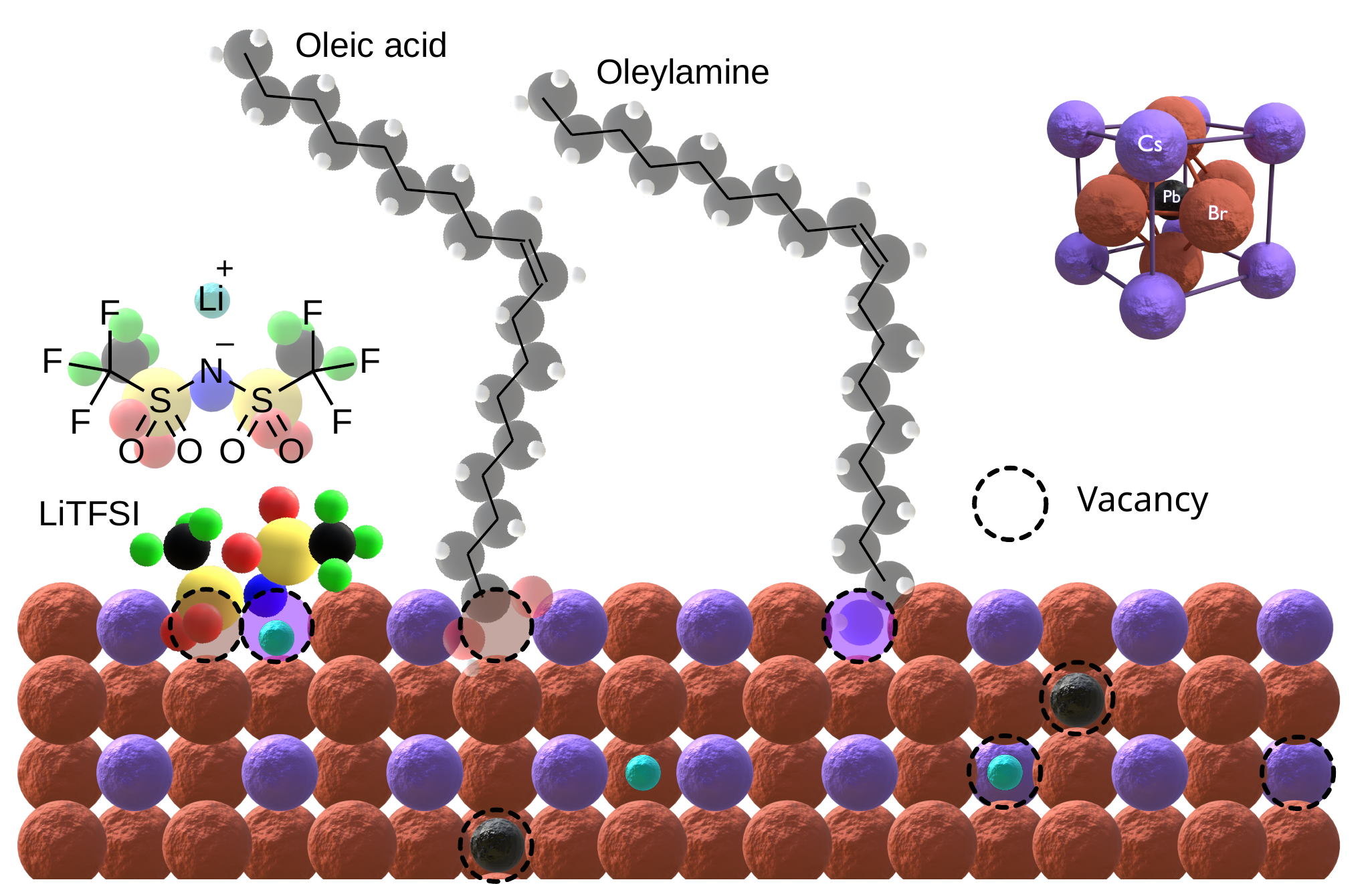}%
		\caption{Schematic \CPB NC surface with possible interaction mechanisms by oleic acid, oleylamine and the organic lithium salt, LiTFSI. Li\tS+ may fill a vacancy or may be intercalated. On the upper right, a full PbBr\ts6 octahedron with adjacent Cs ions is depicted to illustrate the ABX\ts3 crystal structure.}%
		\label{surfPassScheme}%
	\end{figure}

	\paragraph{}{%
		Electroluminescent perovskite light-emitting devices (PeLED) have already been reported in the 1990s and achieved external quantum efficiencies similar to state-of-the-art fluorescent organic LEDs at that time;\cite{Era1994} %
		however, due to severe temperature-induced efficiency drop they have been operated at low temperature only, typically in liquid nitrogen.\cite{Coelle2001} %
		Triggered by their success in photovoltaics, perovskites with the general formula ABX\ts3 (where A is a monovalent organic or inorganic cation, B a bivalent cation – typically Pb\tS{2+} – and X a halide anion) have been “rediscovered” as light emitters less than a decade ago\cite{Tan2014} and have developed into a rapidly progressing LED technology ever since.\cite{Liu2019, Liu2020, Park2019, Ji2021} %
		Narrow emission bands, which are easily tunable by perovskite composition and size in the case of nano-structured materials, paired with high photo- and electroluminescence efficiency over the entire visible and near infrared spectral range make them attractive candidates for next generation displays and lighting. %
		However, despite external quantum efficiencies of PeLEDs being close to organic LEDs, their practical use is still severely limited by insufficient operational lifetimes being on the order of a few 10-100 hours at best.\cite{Woo2021, Kim2021} %
	}%
	\paragraph{}{%	
		In the focus of this work are electroluminescent lead halide perovskite (LHP) nanocrystals (NC), which are, beside 3D bulk and 2D layered perovskites, the third important class of active light emitters for PeLEDs. %
		All three share attractive features like solution processability, ease of bandgap tuning and defect tolerance, however, with potentially higher radiative decay rates and, thus, higher photoluminescence quantum yield in the case of LHP NCs even in the absence of a core-shell structure. %
		Nevertheless, NCs have particularly large surface-area-to-volume ratio, which makes them very sensitive toward surface defects and, thus, strategies to develop defect passivation are particularly important.\cite{Ye2021} %
		Consequently, surface chemistry plays a pivotal role in the optoelectronic properties of LHP NCs. %
	}%
	\paragraph{}{%
		LHP NCs are typically synthesised as colloidal suspensions stabilised by organic ligands, such as oleic acid and oleylamine (see Fig.\,\ref{surfPassScheme}).\cite{Protesescu2015} %
		Despite their high photoluminescence, these pristine “solutions” cannot be used directly for film fabrication and implementation in PeLEDs due to the electrically insulating nature of the ligands. %
		Moreover, non-binding excess ligands, required to stabilise the suspensions, even deteriorate charge injection into the NCs further.\cite{Gomez2017, Kumawat2018, Yan2021} %
		“Washing” the nanocrystals before film deposition removes these ligands partially, however, at the expense of creating a large density of surface defects with the concomitant formation of non-radiative exciton decay channels mainly caused by A- and X-site vacancies, as well as inducing stability issues. %
		To overcome these problems, a variety of different surface passivation strategies have been developed and employed in LHP NCs, including halide salts, strongly binding organic ligands and zwitterionic species.\cite{Li2018} %
		In particular, the use of lithium halides has been shown to increase the photoluminescence yield as well as their stability.\cite{Wu2020} %
		Therein, its passivation is mostly attributed to the excess halides that fill the corresponding vacancies at the NC surface and, thus, reduce non-radiative defects and, at the same time, suppress ion migration via these vacancies. %
		Nevertheless, the role of the Li\tS+ cation itself has largely remained elusive.\cite{Wu2021} %
		In this work, we show that by adding an organic lithium salt, viz. Lithium bis(trifluoromethanesulfonyl)imide (LiTFSI, see Fig.\,\ref{surfPassScheme}), to various LHP NC solutions, with emission colours ranging from deep red, via green to sky blue, their optoelectronic properties are tuned towards higher photoluminescence quantum yield, better hole injection and transport, as well as lower electroluminescence turn-on voltage. %
		Overall, this results in a four- to seven-fold increase of the external quantum efficiency of PeLEDs. %
	}%
	\section{Results and Discussion}%
	\subsection{Surface ligand exchange}%
	\paragraph{}{%
		LHP NCs with different stoichiometry and emission colour, ranging from sky blue (CsPbBr\ts2Cl), via green (\CPB) to red (MAPbBrI\ts2, where MA stands for methylammonium), have been obtained commercially or synthesised as described in the Methods section. %
		The native NC solutions with oleylamine and oleic acid surface ligands (10\,mg/ml solid NC contents in toluene) have subsequently been mixed with equal volumes of LiTFSI solutions in chlorobenzene; see Methods for details. %
		This has resulted in different weight concentrations of LiTFSI relative to the LHP NCs, such as 0.1\wtpct, 1\wtpct, 9\wtpct and 50\wtpct. %
		Thin films have been prepared by spin-coating under inert atmosphere using these LiTFSI:NC mixed solutions, and their properties compared with corresponding thin films made from the native NC solutions. %
		In the following, we focus on green \CPB NCs; blue and red NCs are only discussed in the LED part of this manuscript, but we anticipate that they behave qualitatively similar regarding the effect of LiTFSI on optical, electronic and charge transport behaviour.%
	}%
	\paragraph{}{%
		In general, film formation without and with LiTFSI is found to be very similar (see Fig.\,S4) so that we will not discuss this in detail here. %
		\CPB NCs of size 7-8\,nm arrange in a cubic packing of the NCs on the surface of various kinds of substrates. %
		We do not find a significant difference in NC arrangement and coverage, indicating that the addition of LiTFSI to the NC solutions preserves a certain fraction of the native oleylamine and oleic acid ligands, as schematically shown in Fig.\,\ref{surfPassScheme}. %
		These films were subsequently studied by photoluminescence, photoelectron spectroscopy, density functional theory modelling, and in PeLEDs.%
	}%
	\subsection{Photoluminescence}%
	\begin{figure}[hbt]%
		\begin{minipage}{0.49\linewidth}%
			\begin{overpic}[width=\linewidth]{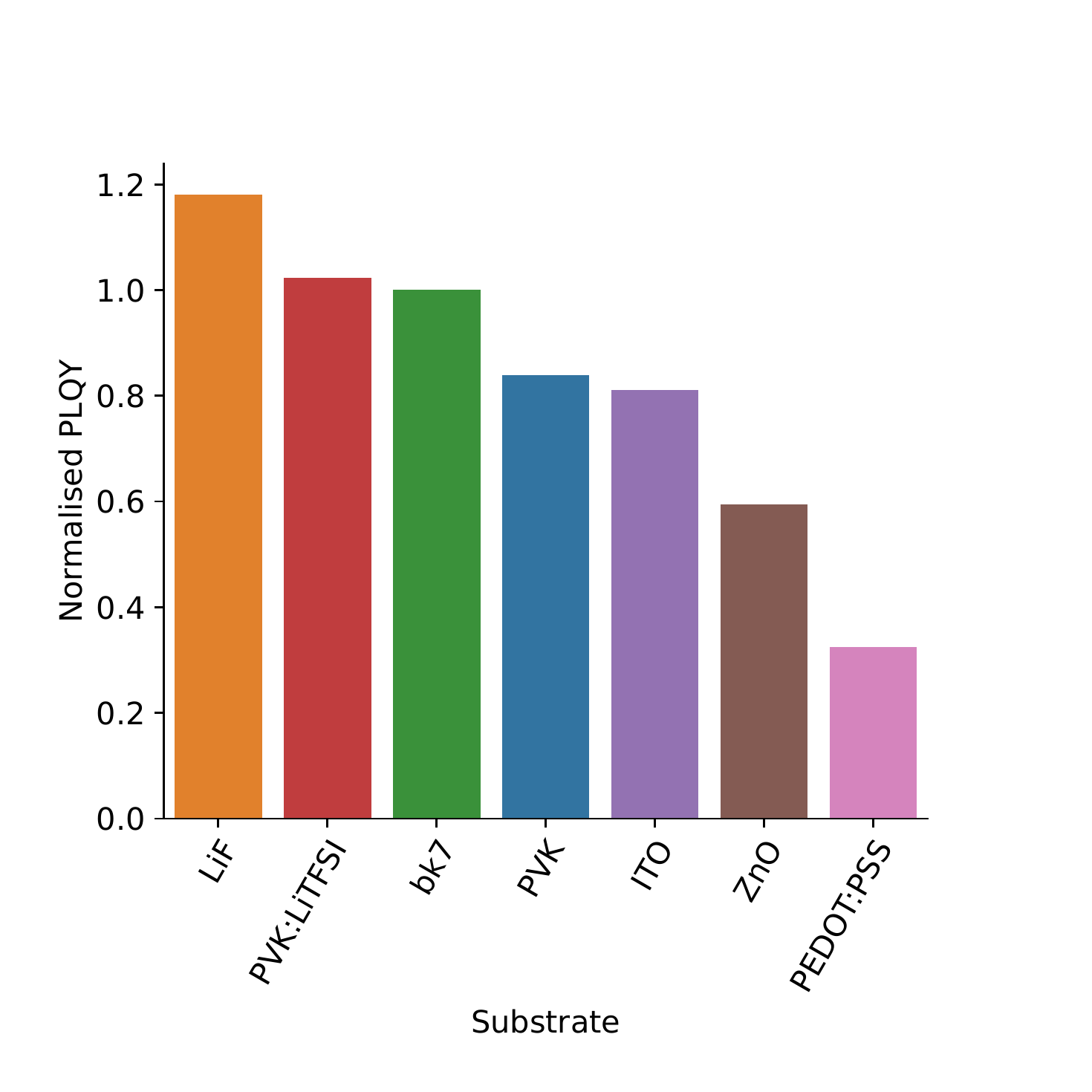}%
				\put (5,95) {\scriptsize a)}%
			\end{overpic}%
		\end{minipage}{}%
		\begin{minipage}{0.49\linewidth}%
			\begin{overpic}[width=\linewidth]{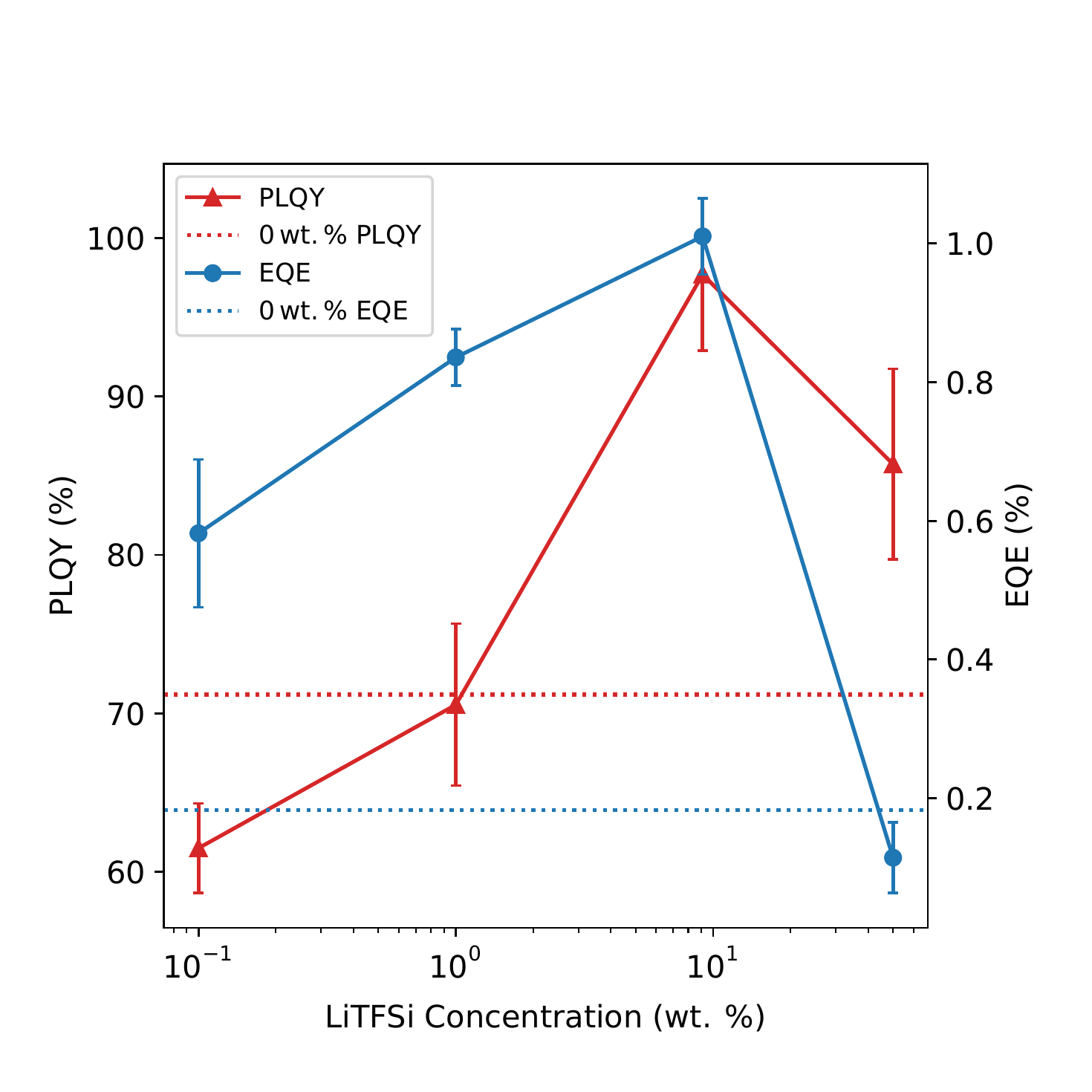}%
				\put (5,95) {\scriptsize b)}%
			\end{overpic}%
		\end{minipage}\\%
		\begin{minipage}{0.49\linewidth}%
			\begin{overpic}[width=\linewidth]{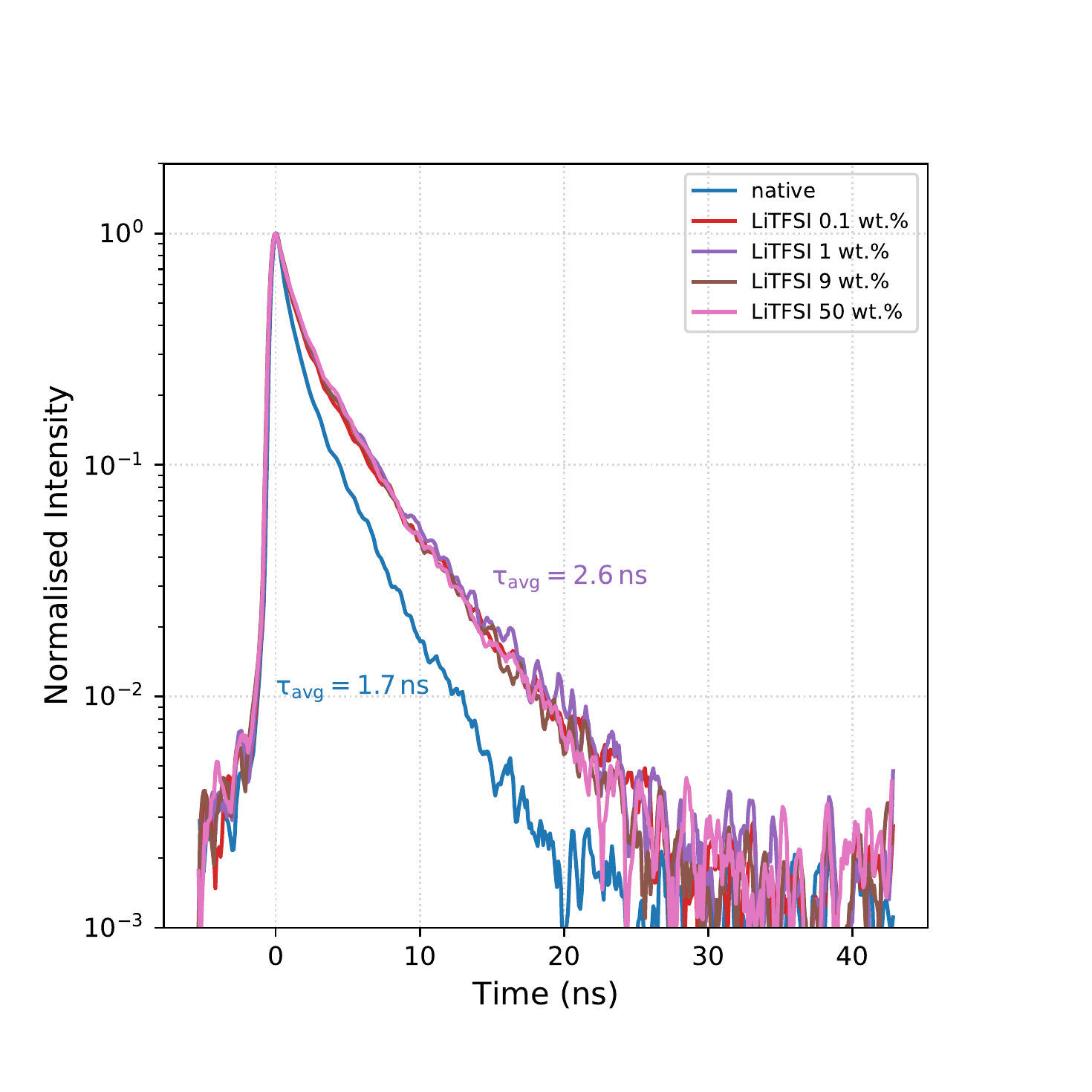}%
				\put (5,95) {\scriptsize c)}%
			\end{overpic}%
		\end{minipage}{}%
		\begin{minipage}{0.49\linewidth}%
			\begin{overpic}[width=\linewidth]{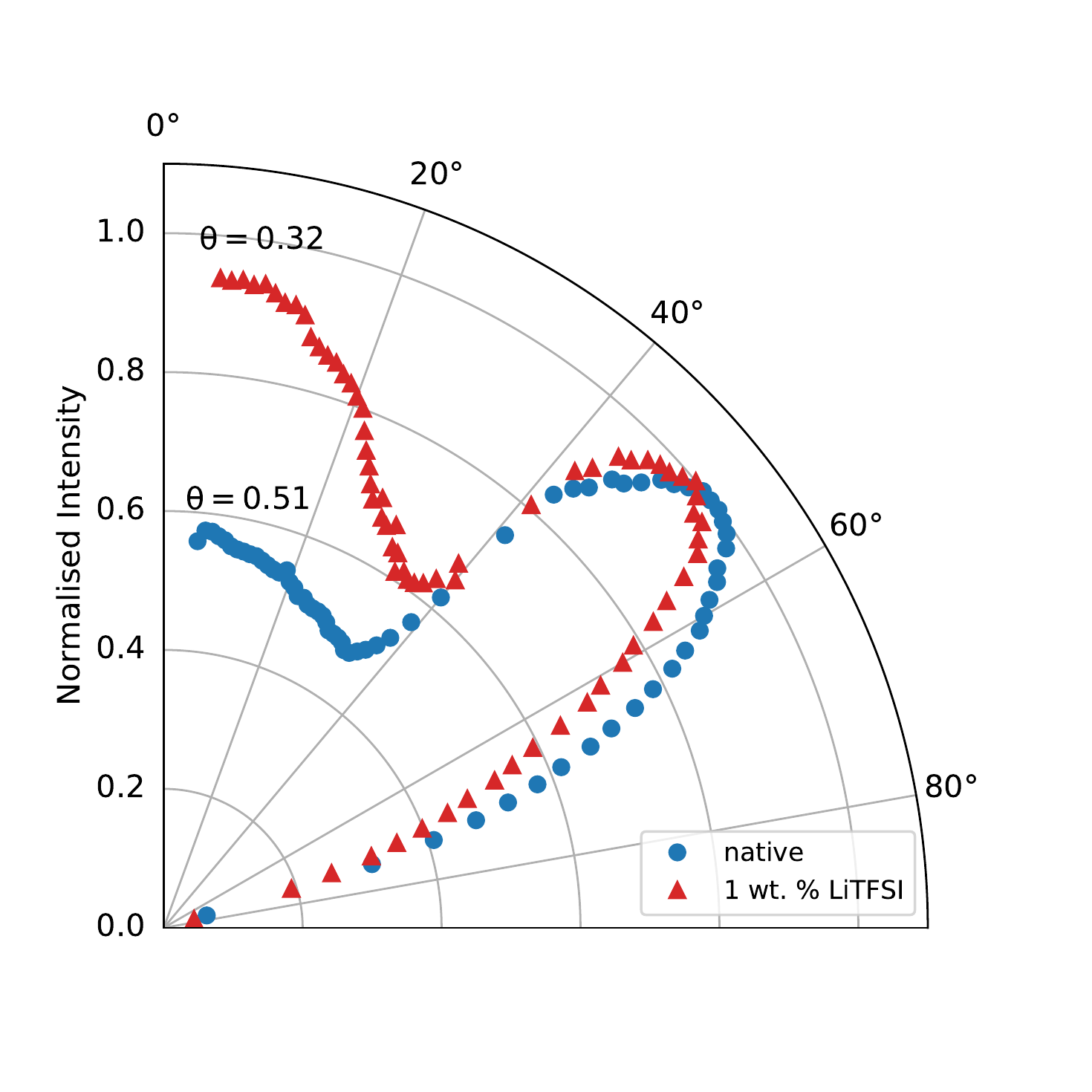}%
				\put (5,95) {\scriptsize d)}%
			\end{overpic}%
		\end{minipage}%
		\caption{a) Normalised PLQY of \CPB NC thin films on various substrates; b) LiTFSI doping concentration dependent PLQY and EQE; c) TRPL decay curves; d) p-polarised ADPL spectra.}%
		\label{PL}%
	\end{figure}%
	\paragraph{}{%
		Photoluminescence quantum yield (PLQY) and lifetime are key indicators for the structural intactness of the LHP NC surfaces, both in solution as well as in thin films. %
		Thereby, not only ligand passivation plays a key role but also their dielectric environment and the substrate on which they are deposited. %
		Fig.\,\ref{PL}\,a shows a comparison of PLQY of native, i.e. oleylamine/oleic acid-capped, \CPB NCs on various substrates relative to glass (bk7). %
		When using highly conductive substrates like metals, ITO, ZnO or PEDOT:PSS the PLQY is significantly lowered compared with glass. %
		For PEDOT:PSS the quantum yield reduction can be ascribed to residual water content which may degrade the LHP NC significantly and quench luminescence.\cite{Nguyen2004, Peng2017} %
		The reduction for ZnO and ITO however originates from a different mechanism; e.g. their high carrier densities may enable exciton-carrier quenching or the oxides could lead to degradation of the LHPs.\cite{Olthof2021} %
		And even in the case of the organic polymer Polyvinylcarbazole (PVK), used as hole transporting layer (HTL) in PeLEDs, the PLQY is slightly lowered with respect to glass. %
		However, by introducing LiTFSI as dopant in PVK the impairment is overcome. %
		LiTFSI as dopant in HTLs is reported to increase their electrical conductivity rendering this effect rather unexpected.\cite{Kim2019} %
		Another Li-based substrate treatment is a thin film of LiF evaporated on glass before spin-coating of the LHP NCs, which proves to be even more beneficial on the PLQY (see Fig.\,\ref{PL}\,a). %
		A positive effect of an LiF interlayer has already been observed for light emission from bulk perovskites. %
		Bigger grain sizes and reduced pinholes in the film were identified as the main cause for a better optical performance.\cite{Zhao2020} %
		For LHP NCs the grain sizes are predefined during synthesis and not expected to change upon spin-coating on a substrate.\cite{Protesescu2015} %
		Consequently, it is reasonable to suspect a different mechanism for the PLQY enhancement.%
	}%
	\paragraph{}{%  
		This raises the question, whether the nature of the substrate alone affects the PLQY, or if there is interdiffusion from the underlying material into the LHP NC thin film. %
		Thus, to study the influence of LiTFSI on the \CPB  NCs, we directly add the Li salt to the solution prior to spin-coating on glass. %
		Even though a more polar solvent, viz. chlorobenzene, is needed to mix the two substances, an increase in PLQY can be observed, both, in solution and as a thin film. %
		Fig.\,\ref{PL}\,b shows the PLQYs of thin-films fabricated with different LiTFSI mass percentage mixing ratios. %
		At very low concentrations, the PLQY is slightly decreased with respect to the pristine (0\wtpct) LHP NC film, which has a PLQY of about 70\,\%. %
		In contrast to that, at concentrations higher than 1\wtpct, the PLQY is significantly increased reaching near unity at 9\wtpct, before it decreases again for a 1:1 mixture of both components. %
		The PLQY impairment at low concentrations is comparable to a solvent mixture of toluene and chlorobenzene (but no LiTFSI), which may imply that some of the NCs are irreversibly degraded by the polar solvent before this detrimental effect is outweighted at higher LiTFSI content.\cite{Sun2021}%
		}%
	\paragraph{}{%
		Since we suspect a surface trap passivation effect by the LiTFSI, time-resolved photoluminescence (TRPL) has been measured on the same films (Fig.\,\ref{PL}\,c). %
		As detailed in the Supporting Information (Fig.\,S1\,\&\,2), the PL transients show a double-exponential decay with a fast initial lifetime of about 1\,ns and a slower one with several nanoseconds.
		Weighting the lifetimes with their relative amplitudes, one obtains an average lifetime, as explained in the context of Fig.\,S1\,\&\,2.
		For the native sample $\tau$\textsubscript{avg}\,=\,1.7\,ns is obtained, whereas all the LiTFSI doped ones have similar $\tau$\textsubscript{avg}\,=\,2.6\,ns, with only minor variation $\Delta\tau$\textsubscript{avg}\,=\,0.1\,ns (see Fig\,\ref{PL}\,c), which is surprising in view of the observed changes in the PLQY over the same concentration range. %
		However, as already discussed, the PLQY is reduced if there are solvent-degraded LHP NCs in the film, caused by their optically passive absorption. %
		On the other hand, the PL lifetimes do not necessarily have to be affected by the degraded LHP NC if they do not emit any light and also do not interact with the optically active ones.%
	}%
	\paragraph{}{%
		As the double-exponential fits on the intensity decay (see Fig.\,S1\,\&\,2) reveal, the individual PL lifetimes are concentration independent (just like the average, $\tau$\textsubscript{avg}). %
		However, a comparison of the amplitudes exhibits a shift of weights towards the longer lifetime with increasing LiTFSI concentration. %
		Together with the PLQY measurements, this implies that within the short lifetime there are more non-radiative decay processes. %
		Furthermore, this could indicate that trap-assisted recombination is suppressed by the presence of LiTFSI. %
		There have been several reports about surface trap passivation featuring various halide salts. %
		They agree on a passivation mechanism by halide abundance, which has also been reported by groups using lithium-free halide salts.\cite{Dutta2019,Akkerman2015,DiStasio2017,Woo2017} %
		However, in our case the halide-free LiTFSI seems to have a similar effect. %
	}%
	\paragraph{}{%
		In addition to the changes in PLQY and PL lifetime, we observe a change of the radiation pattern of LHP NC films upon addition of LiTFSI. %
		Fig.\,\ref{PL}\,d shows the parallelly polarised (p-pol.) angular dependent photoluminescence (ADPL) of a native and a 1\wtpct LiTFSI-treated \CPB -NC thin film. %
		As studied in great detail for organic light emitters\cite{Schmidt2017} as well as for some LHP NCs,\cite{Jurow2019, Jagielski2020} such radiation patterns contain information about the average orientation of the transition dipole moments (TDM) of the electronically active optical transition from the excited state to the ground state.
		This parameter is key to understanding and improving light outcoupling from thin-film light-emitting structures, such as organic LEDs\cite{Schmidt2017} or PeLEDs.\cite{Morgenstern2020} %
		In particular, the orientation parameter $\theta$ (which is defined as the fraction of power emitted by vertical TDMs) directly indicates the degree of anisotropy of the TDM orientation distribution, with $\theta = 0.33$ being the isotropic case, and larger (smaller) values indicating more vertical (horizontal) TDM orientation.
		By comparing native \CPB and LiTFSI-treated NCs we notice a significant change of the TDM orientation.
		After fitting with an appropriate dipole model (details in the SI) a TDM orientation parameter of $\theta$\textsubscript{native} = 0.51 is obtained, proving more vertical TDM orientation for the native NCs, in contrast to $\theta$\textsubscript{LiTFSI} = 0.32 being close to isotropic for the 1\wtpct LiTFSI-treated NCs. %
		This means that the unfavourable vertical TDM orientation of the native NC film is changed to an isotropic emission profile corresponding to the cubic structure of LHP. %
		A similar behaviour has also been found by an Al\textsubscript{2}O\textsubscript{3} overcoating of LHP nanoplatelets.\cite{Jurow2017,Jurow2019} %
		Thus, we suppose that LiTFSI acts as a dielectric layer to screen the asymmetry in the electrostatics between the substrate and the free surface of the NCs.%
	}%
	\subsection{Energy Level Alignment}%
	\begin{figure}[htb]%
		\begin{minipage}{0.6\linewidth}%
			\begin{overpic}[width=\linewidth]{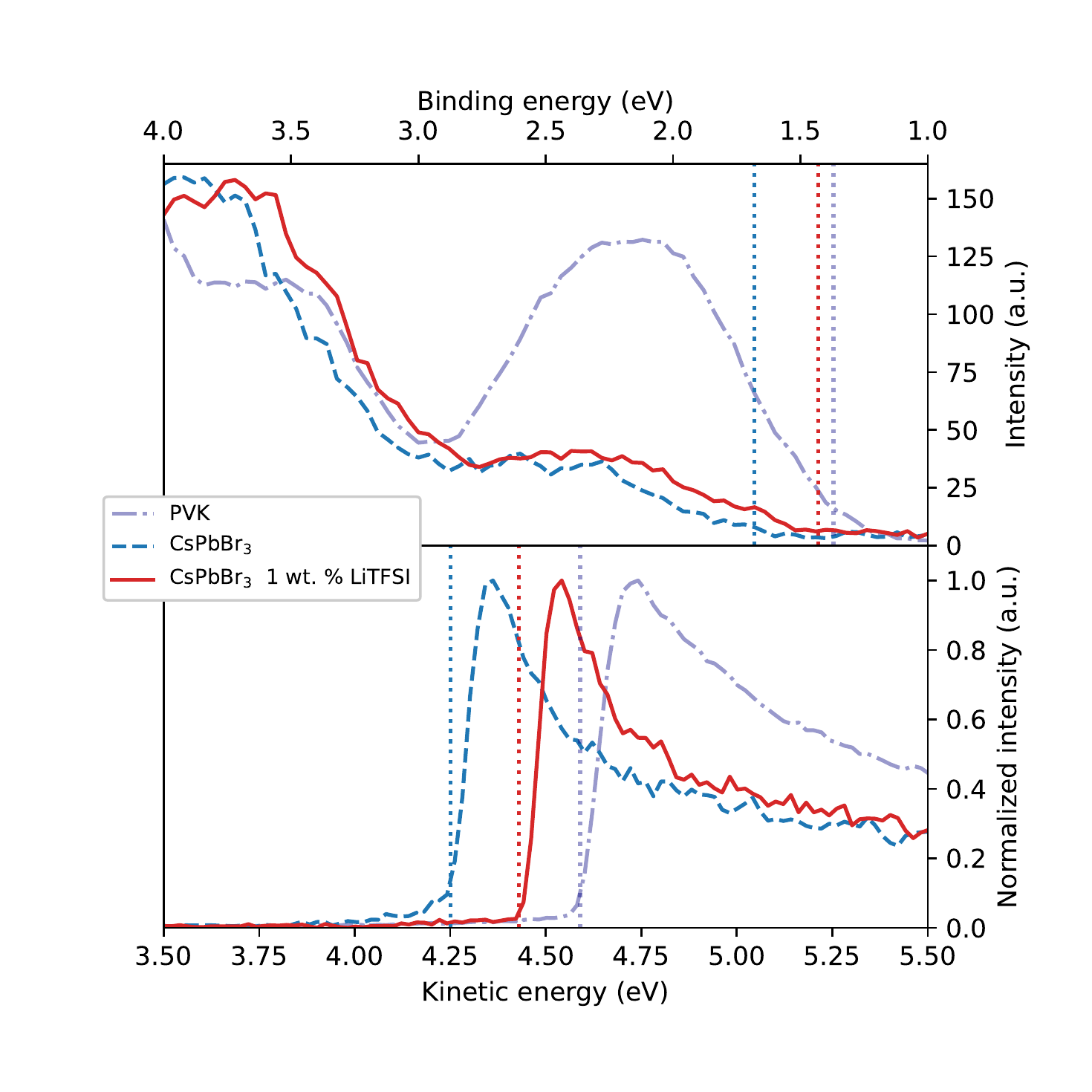}%
				\put (5,95) {\scriptsize a)}%
			\end{overpic}%
		\end{minipage}\\%
		\begin{minipage}{0.6\linewidth}%
			\begin{overpic}[width=\linewidth]{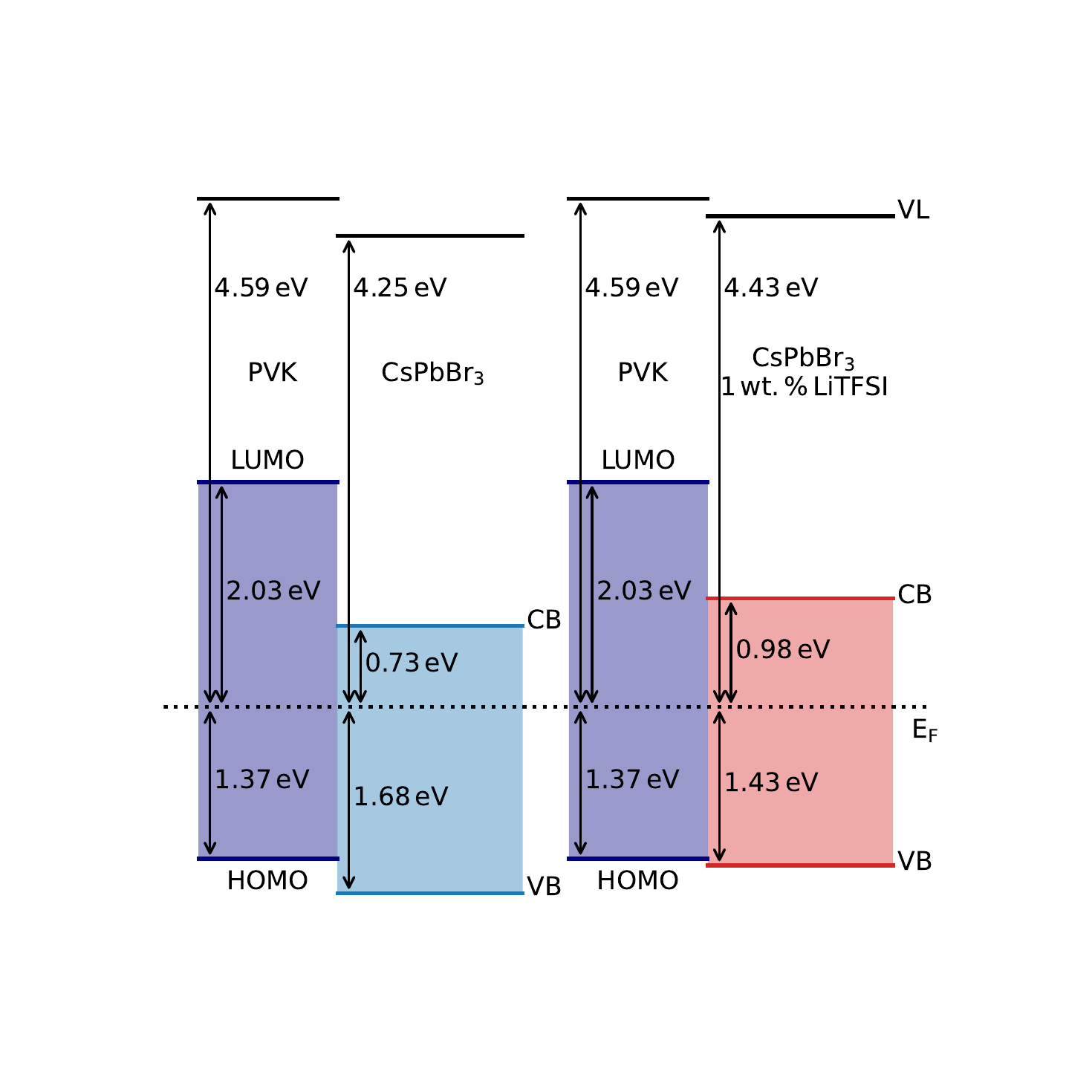}%
				\put (5,95) {\scriptsize b)}%
			\end{overpic}%
		\end{minipage}\\%
		\caption{a) UPS spectra of PVK and PVK/\CPB -NC films with and without addition of LiTFSI. Top panel: valence band spectra; bottom panel: secondary electron cutoff region. b) Energy level diagram at the PVK/\CPB interface with and without addition of LiTFSI. VL and E\ts{F} refer to vacuum level and Fermi level (set at 0\,eV binding energy). Bandgaps are taken from the optical gap for the perovskite and from literature for PVK.\cite{Tavares2021}}%
		\label{UPS}%
	\end{figure}%
	\begin{figure}[htb]%
		\begin{minipage}{0.6\linewidth}%
			\includegraphics[width=\linewidth]{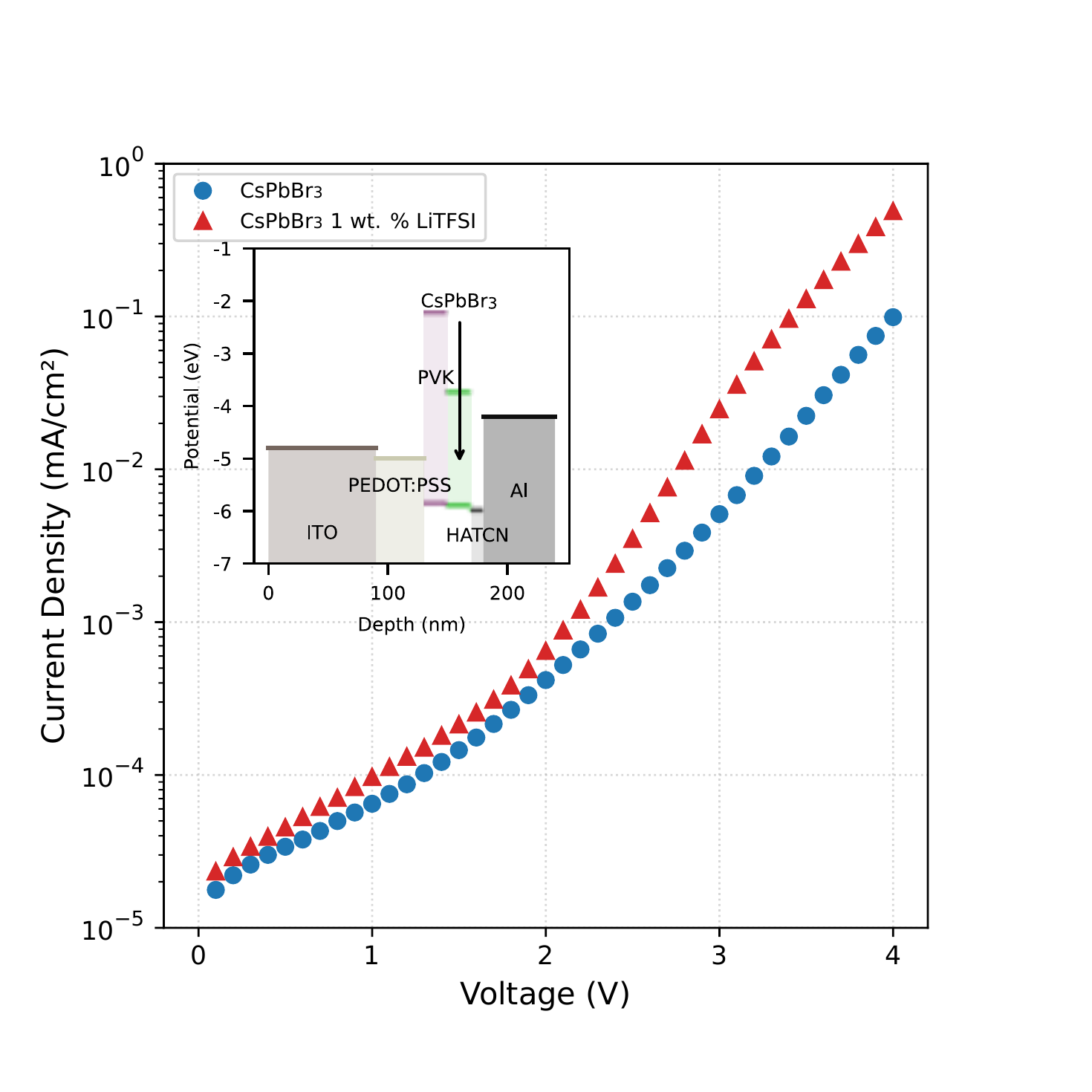}%
		\end{minipage}%
		\caption{Hole-only j-V curve with the schematic device stack as inset.}%
		\label{holeOnly}%
	\end{figure}%
	\paragraph{}{%
		In order to elucidate the impact of LiTFSI on the electronic properties of LHP NCs, ultraviolet and X-ray photoelectron spectroscopy (UPS and XPS) measurements have been conducted on the PVK/\CPB -NC and PVK/\CPB -NC:LiTFSI films, respectively. %
		As shown in Fig.\,\ref{UPS}\,a by the valence region and the secondary electron cut-off spectra, the PVK film exhibits an initial work function of 4.59\,eV and the highest occupied molecular orbital (HOMO) level is found at 1.37\,eV binding energy (with respect to E\ts{F}). %
		Upon deposition of \CPB -NCs, the work function decreases to 4.25\,eV, likely due to the formation of band bending and/or an interface dipole at the buried interface.\cite{Olthof2017} %
		The valence band (VB) onset of \CPB is then extrapolated at 1.68\,eV with respect to E\ts{F}. 
		Given the energy gap of \CPB of about 2.4\,eV (515\,nm), this shows that E\ts{F} is located above mid-gap which is due to a strong n-type character on the surface of the NCs. %
		Additional surface photovoltage measurements under white light illumination shows no shifts of \CPB -NC energy levels, indicating a flat band condition through the LHP NC layer.\cite{Zu2019acsami} %
		Hence, the surface energy levels is then expected to also reflect the electronic properties within LHP NC thin-film. %
		With the addition of LiTFSI to the \CPB -NCs, a shift of the \CPB -NC valence band towards lower binding energy by 0.25\,eV is observed, which is accompanied by an increase of sample work function, leading to a decrease of VB onset to 1.43\,eV with respect to E\ts{F}. %
		A similar shift of the core levels has been observed in XPS (see Fig.\,S5). 
		Such a rigid shift of all \CPB -NC energy levels distinctly demonstrates a p-doping effect by the addition of LiTFSI. %
		It is worth mentioning that the use of substantially attenuated UV flux (attenuation of more than 100 times compared to the standard helium discharge lamp) is required for the UPS measurement, as a high UV flux is found to cause irreversible changes of the electronic structure. %
		However, this leads to insufficient signal-to-noise ratio at the top VB region, which refrains us from extracting the LHP NC valence band onset on a logarithmic intensity scale of the photoelectron signal, as is known to accurately infer the band edge position of perovskite films due to the low density of states at the top of the valence band.\cite{Endres2016, Zu2019} %
		However, the shift of the electronic levels is not affected by this procedure. %
		The energy level diagram of the PVK/\CPB -NC stack is shown in Fig.\,\ref{UPS}\,b. %
		It can be clearly seen that PVK/\CPB -NC interface initially exhibits a large energy barrier of ca. 0.31\,eV for hole injection. %
		Due to the p-doping effect by addition of LiTFSI, such energy barrier is reduced to 0.06\,eV with \CPB -NC VB edge shifting closer to the HOMO level of PVK. %
		With such reduction, the hole injection can be significantly improved.%
	}%
	\paragraph{}{%
		The reduced hole injection barrier is manifested in the electrical transport behaviour as well. %
		As expected a single carrier hole-only device (layer stack shown in the inset of Fig.\,\ref{holeOnly}) with 1\% LiTFSI doping shows a significant increase, respectively doubles the current density when operated at 4\,V (see Fig.\,\ref{holeOnly}). %
		We note that this current enhancement is not caused by a change of the layer morphology (including its thickness). %
		As shown in the Supporting Information (Fig.\,S4), both native and LiTFSI-treated \CPB -NC films have similar morphology with a partially closed uppermost NC monolayer on top of fully closed layer(s) underneath. %
		It rather seems that the LiTFSI treatment improves the film morphology so that the increase of current cannot originate from insufficient NC coverage. %
		Further, one can also expect to obtain different electronic coupling among the \CPB -NCs by replacing (part of) the insulating oleylamine/oleic acid ligands by LiTFSI so that an improved bulk charge transport might also contribute to the observed current enhancement.%
	}%
	\subsection{Density Functional Theory Modelling}%
	\begin{figure}[htb]%
		\begin{minipage}{\linewidth}%
			\includegraphics[width=\linewidth]{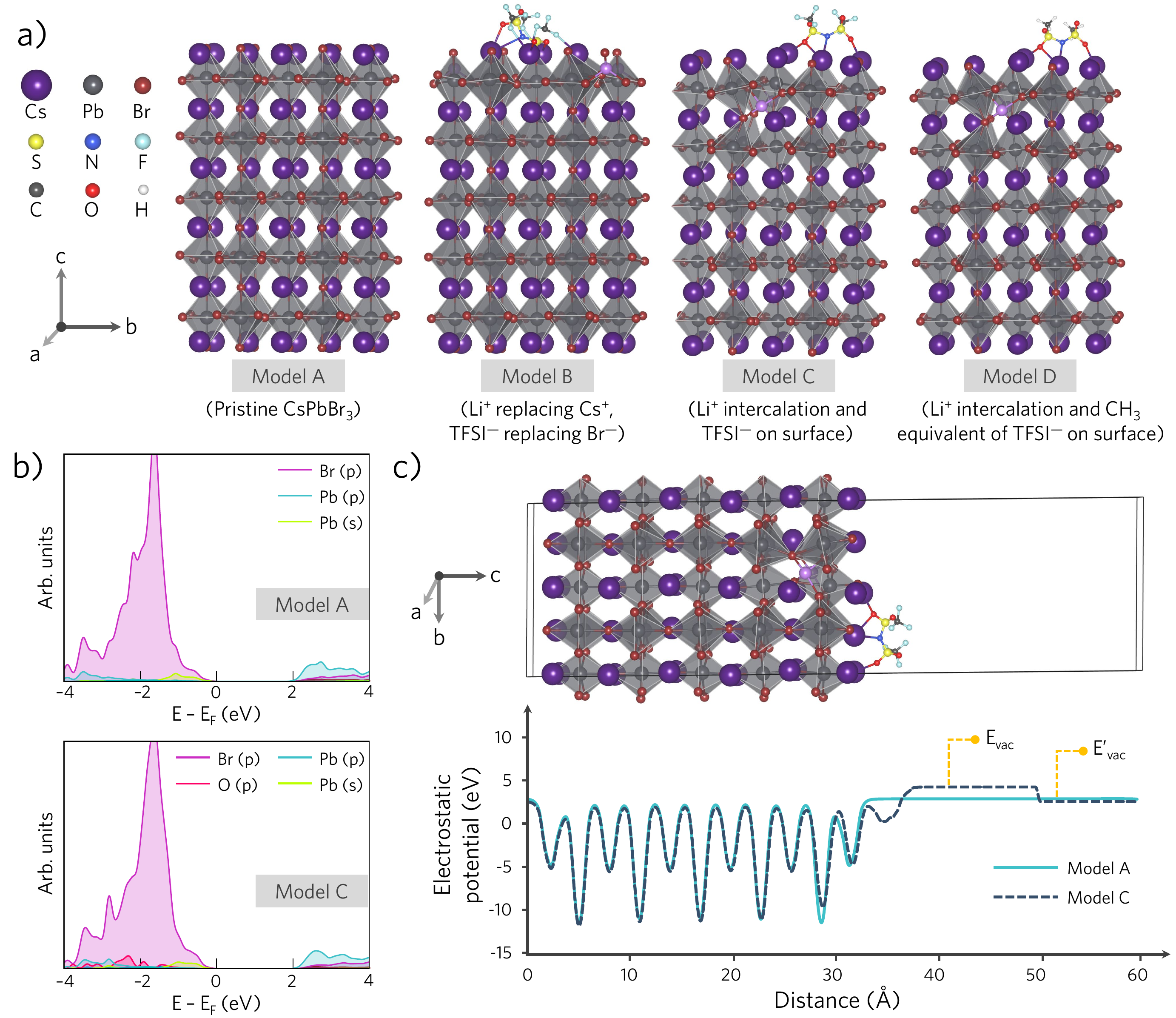}%
		\end{minipage}%
		\caption{a) Slab models of orthorhombic \CPB (1×2×3 supercell; 5 PbI-layers) exposing the CsBr-terminated surfaces, b) projected density of states plots for model A and C (E\ts{F}\,=\,Fermi energy) c) planar average electrostatic potential for determining vacuum energy level, E\ts{vac}. In the diagram, E\ts{vac} corresponds to the (near) vacuum level of the TFSI-deposited surface and E$^\prime$\ts{vac} represents the undoped (far) vacuum level.}%
		\label{DFT}%
	\end{figure}%
	\paragraph{}{%
        Density functional theory (DFT) modelling of four slab models was performed (see Supporting Information for computational details) to understand the structural changes upon LiTFSI treatment and its consequent impact on the electronic structure of the \CPB NCs. As shown in Fig.\,\ref{DFT}\,a, model A consists of pristine \CPB, while in model B, a Li\tS+ ion replaces a Cs\tS+ ion together with a TFSI\tS{\textminus} replacing a Br\tS{\textminus} ion over the surface. Model C considers intercalation of a Li\tS+ ion, with one TFSI\tS{\textminus} placed over the CsBr-terminated surface, while model D is a system analogous to model C, but with the two –CF\ts3 groups of the TFSI\tS{\textminus} ion having been replaced with two –CH\ts3 groups.\\%
        Even though model D is not relevant experimentally, we explored this system computationally to disentangle the effect of C–F bond polarity on work function (WF), by closely comparing it with its C–H analogue.\\%
        We found that both pristine and LiTFSI-doped systems exhibit a very similar density of states (Fig.\,\ref{DFT}\,b), %
        and the orbital contributions from the TFSI\tS{\textminus} ions are quite deep-lying, appearing only near -2.0\,eV. %
        On the other hand, the TFSI\tS{\textminus} ligands were found to alter the surface dipole moment of the particle significantly, %
        causing a substantial shift in the vacuum level (Fig.\,\ref{DFT}\,c). %
        Consequently, the mere presence of a surface TFSI\tS{\textminus} ion in model B replacing a Br\tS{\textminus} ion increased the work function of the NC from 4.51\,eV in pristine model A to 5.54\,eV in model B. %
        The vacuum level-shift is even more prominent in model C, partly attributable to the fact that the dipoles caused by the surface TFSI\tS{\textminus} ions are now exclusively outside the inorganic core, %
        as opposed to model B, where the O-atom from TFSI binds to a Pb\tS{2+} ion from the core and hence, the dipole is partly compensated. %
        In model D, we show that when the highly polar C–F bonds in TFSI are replaced with C–H bonds, %
        the vacuum level shift is drastically reduced (E\ts{vac}\,=\,4.23\,eV and WF\,=\,6.20\,eV for –CF\ts3 groups in model C vs. E\ts{vac}\,=\,3.83\,eV and WF\,=\,5.69\,eV for –CH\ts3 groups in model D). %
        We recognize that the computed change in the work function is rather substantial compared to the experiment (UPS measured hole stabilisation being around 0.25\,eV, Fig.\,\ref{UPS}\,b). We attribute this discrepancy to (i) a higher TFSI\tS{\textminus} coverage in the model system compared to the experiment and (ii) the fact that in the slab model only one surface is asymmetrically covered with organic ligand as opposed to all six-side coverage in an actual NC.\\%
        Besides the change in work function, the topotactic intercalation of the Li\tS+ ion, %
        irrespective of whether in a tetrahedral or an octahedral site, %
        has been demonstrated to increase interaction with neighbouring halide ions (see Li–Br bonding in addition to the usual Cs–Br bonds in model C and D in Fig.\,\ref{DFT}\,a),\cite{Wu2021} %
        and thus the activation barrier for halide ion migration is expected to increase.\cite{Cao2018} %
        We anticipate a similar effect of the Li\tS+ ion on preventing ion migration and improving the structural integrity of the LiTFSI-treated \CPB particles. %
        Finally, we speculate that the availability of electronegative O-atoms from the TFSI\tS{\textminus} ligand, %
        which is quite mobile over the surface, would also coordinate to Pb-atoms, as seen in model B, %
        and contribute towards preventing detrimental Pb\tS{2+} to Pb\tS0 reduction over the surface.\cite{An2018}%
	}%
	\subsection{Perovskite Light-emitting Diodes}%
	\begin{figure}[htb]%
		\begin{minipage}{0.49\linewidth}%
			\begin{overpic}[width=\linewidth]{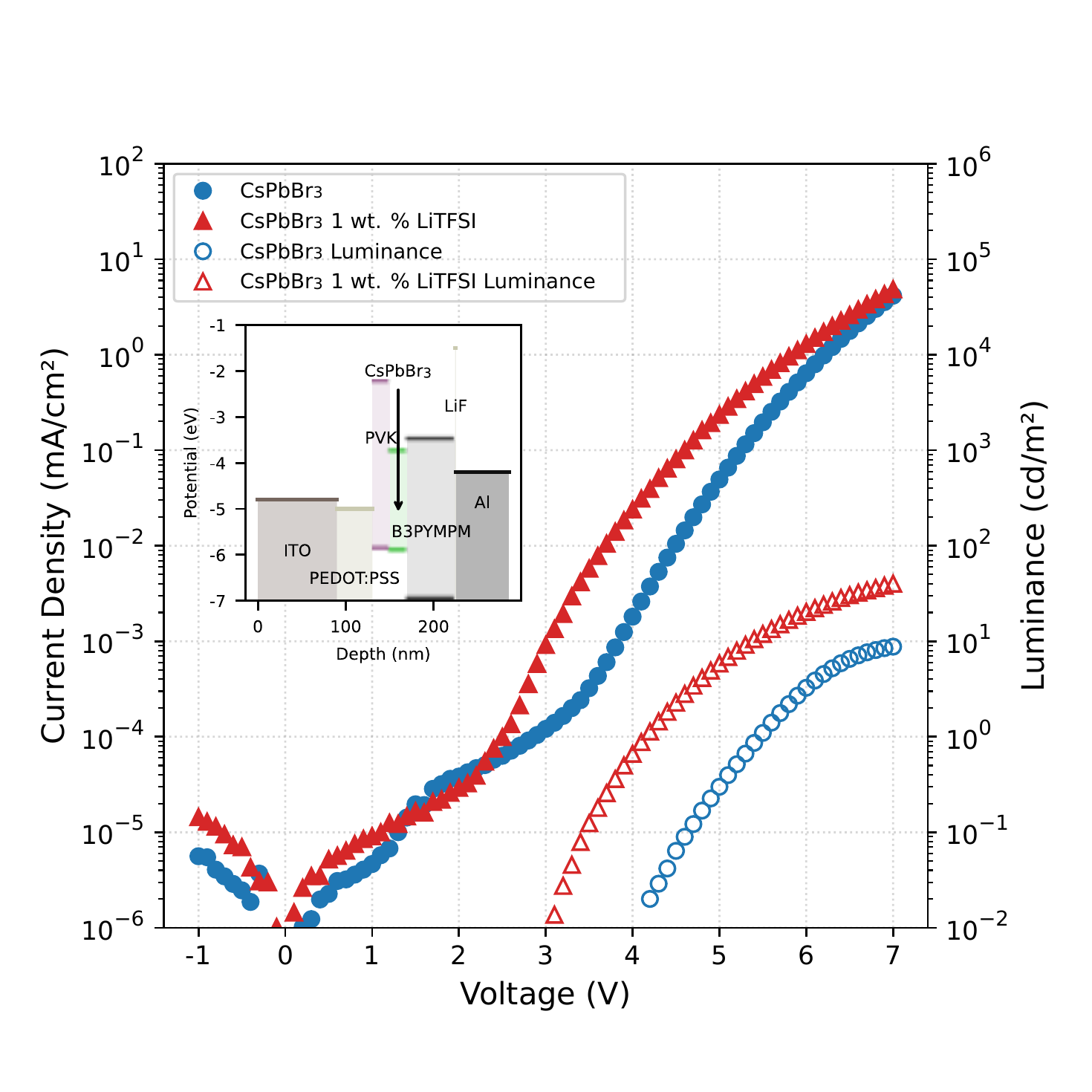}%
				\put (5,95) {\scriptsize a)}%
			\end{overpic}%
		\end{minipage}{}%
		\begin{minipage}{0.49\linewidth}%
			\begin{overpic}[width=\linewidth]{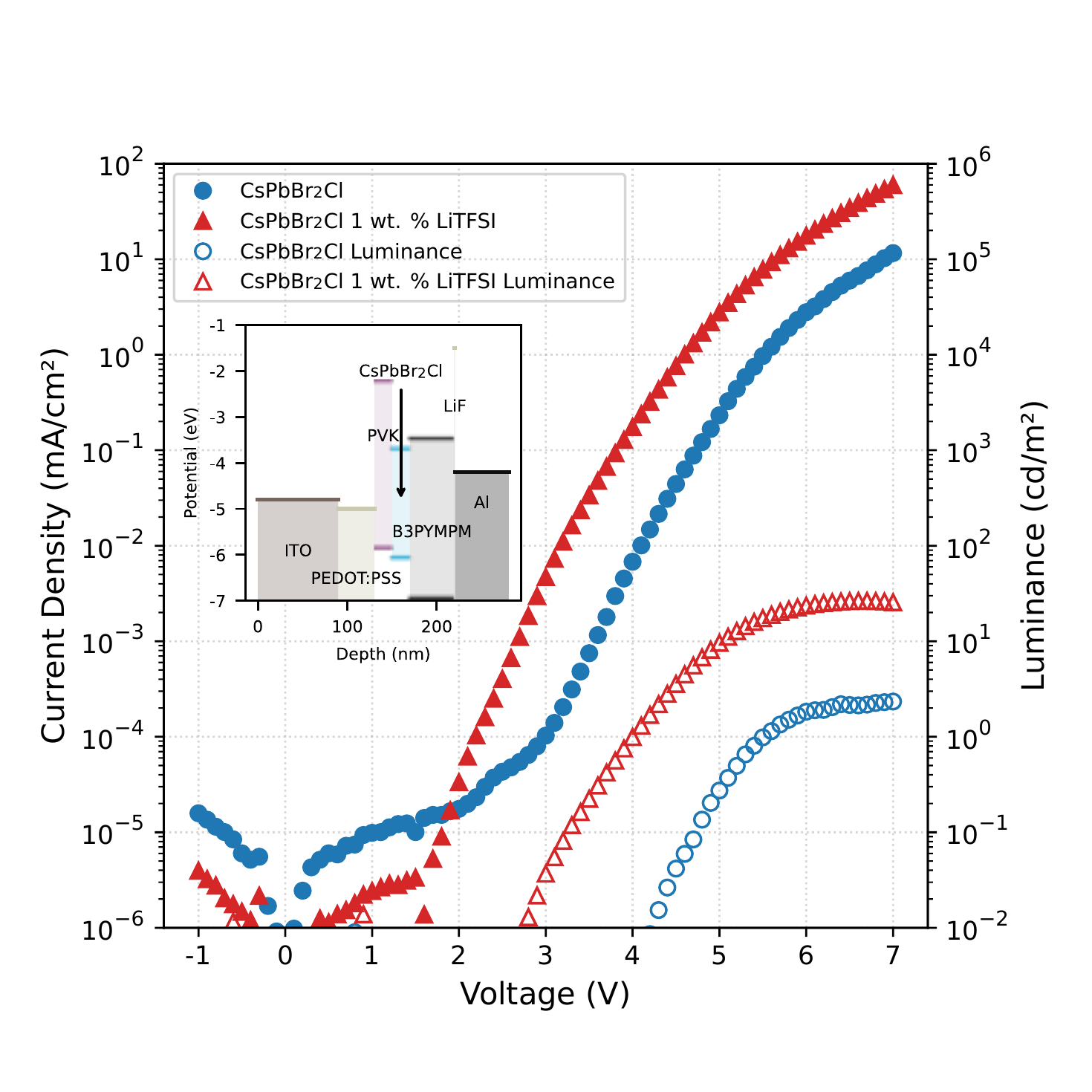}%
				\put (5,95) {\scriptsize b)}%
			\end{overpic}%
		\end{minipage}\\%
		\begin{minipage}{0.49\linewidth}%
			\begin{overpic}[width=\linewidth]{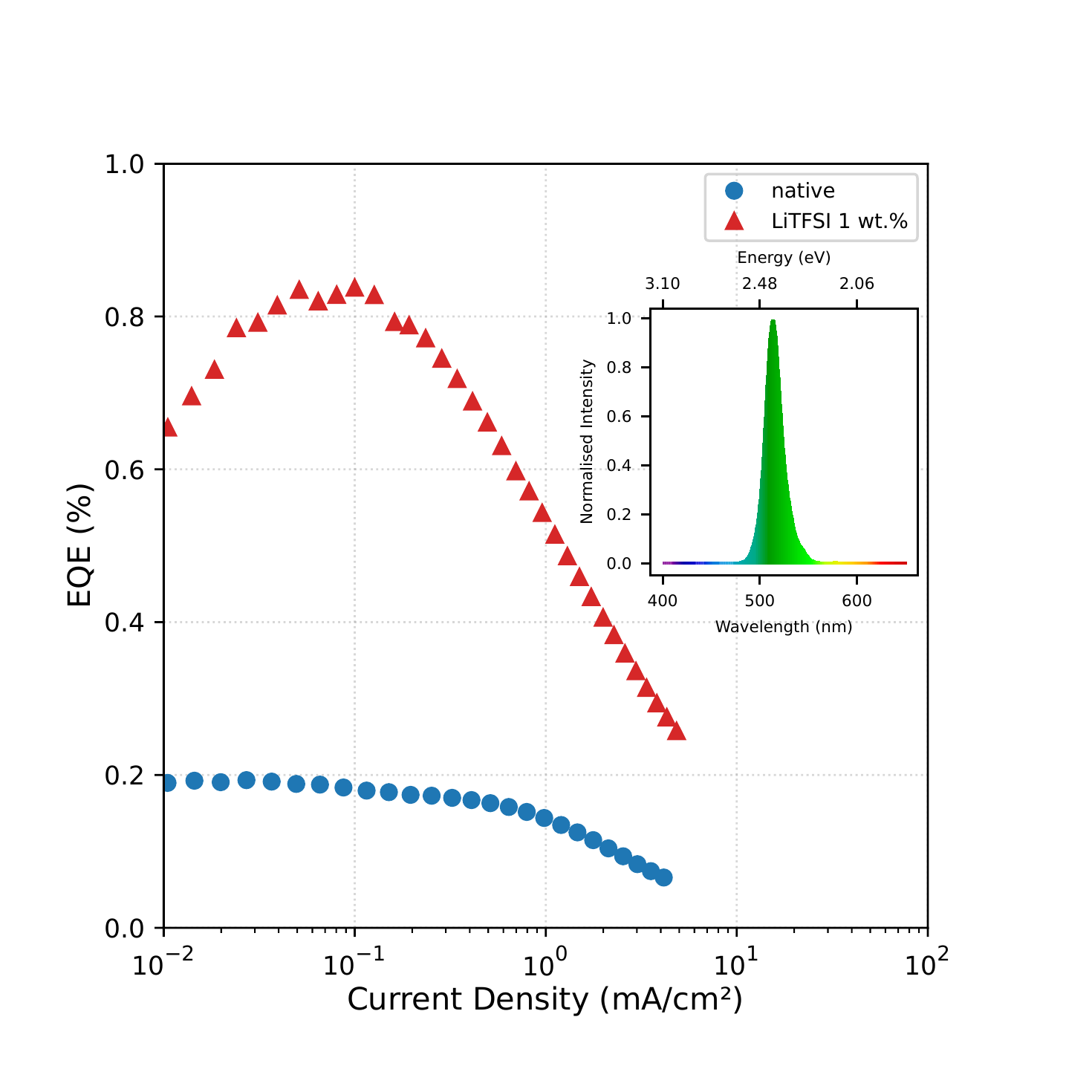}%
				\put (5,95) {\scriptsize c)}%
			\end{overpic}%
		\end{minipage}{}%
		\begin{minipage}{0.49\linewidth}%
			\begin{overpic}[width=\linewidth]{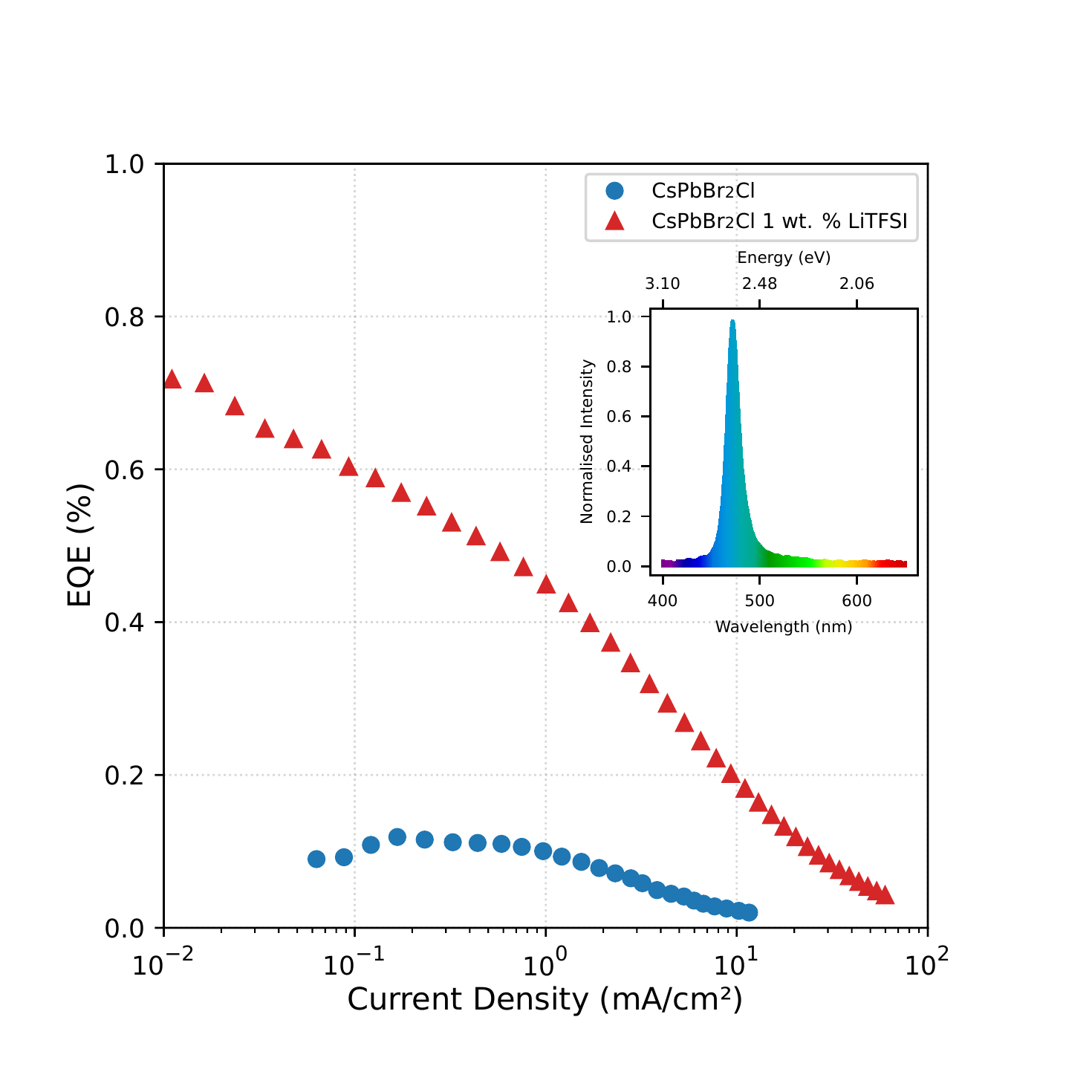}%
				\put (5,95) {\scriptsize d)}%
			\end{overpic}%
		\end{minipage}%
		\caption{j-V-L curves of a) \CPB and b) CsPbBr\ts2Cl PeLEDs with and without LiTFSI-doping; c) \& d) show the corresponding EQE vs current density curves. The insets depict their stack structures and their emission spectra.}%
		\label{jvl}%
	\end{figure}%
	\paragraph{}{%
		Finally, we have employed native and LiTFSI-treated \CPB NCs in PeLEDs, where PVK is used as polymeric hole transport layer and B3PYMPM as electron transport layer, followed by a LiF/Al cathode as depicted in the inset of Fig.\,\ref{jvl}\,a, %
		the corresponding current-voltage-luminance characteristics is also shown in Fig.\,\ref{jvl}\,a. %
		The native \CPB -NCs show two onsets for the current, with the first occurring slightly above 1\,V which can be assigned to electrons due to the n-type position of the Fermi level; %
		the second onset with a much steeper increase of current is located at about 3.5\,V and would accordingly correspond to the onset of hole injection. %
		This is supported by the fact that detectable luminance ($> 10^{-2}\,\mathrm{cd/m^2}$) occurs not before the applied voltage exceeds 4\,V.%
	}%
	\paragraph{}{%
		For the 1\wtpct LiTFSI-doped NCs, the double-step feature in the current disappears; rather, there is a steep increase starting above 2\,V and luminance is detected already at 3\,V. %
		With further increasing voltage, both current and luminance increase much steeper in the LiTFSI-doped PeLED as compared to the untreated one. %
		Finally, at 7\,V both devices reach the same current, but the luminance in the LiTFSI-doped PeLED stays several times above the native one, indicating higher external quantum efficiency (EQE). %
		This is shown in Fig.\,\ref{jvl}\,c where the maximum EQE increases by more than a factor of 4 from 0.2\,\% to above 0.8\,\%. %
		We also note that both, native and LiTFSI-treated \CPB, PeLEDs have the same narrow green electroluminescence spectrum shown in the inset of Fig.\,\ref{jvl}\,c.%
	}%
	\paragraph{}{%
		As a proof of concept, we also applied LiTFSI doping to sky-blue emitting CsPbBr\ts2Cl-NCs (results shown in Fig.\,\ref{jvl}\,b,d). %
		In this case, the reduction of the turn-on voltages for current and luminance induced by LiTFSI doping is even bigger, as detectable luminance sets in already below 3\,V. %
		The EQE is enhanced by a factor 7 from 0.1\,\% to 0.7\,\%, proving an even stronger effect of LiTFSI doping in the wider band-gap, sky-blue PeLEDs.%
	}%
	\paragraph{}{%
		Furthermore, a red methylammonium-PbBrI\ts2-NC PeLED can also be improved by LiTFSI, without an evident current-density increase (see Fig.\,S7). %
		In this case, the red NC emitter does not face the same hole injection barriers as its blue and green counterparts, such that an increase in hole-injection is not expected to play a major role. %
		Nevertheless, the device EQE is boosted by a similar factor through the LiTFSI treatment.%
	}%	
	\subsection{Discussion}%
	\paragraph{}{%
		The overall increase in PeLED performance with various LHP species implies a very versatile mechanism. %
		The increase in PLQY and exciton lifetime with the LiTFSI concentration suggests a surface passivation mechanism. %
		This has already been reported with lithium halides, where the lithium ions as well as the halides have shown surface trap passivation.\cite{Wu2020,Wu2021,Fang2018}
		However, the influence of Li\tS+ has not been completely disentangled from the halide ones.
		Here the anion is not a halide, but the organic TFSI\tS{\textminus} ion, proving that a halide-free enhancement is possible. %
		Still, the TFSI may not be a passive agent on the surface of LHP NCs. %
		As sulfonic moeities have been reported to fill halide vacancies, a similar interaction from TFSI may be thinkable.\cite{Ye2021} % 
		Ultimately, a similar passivation as reported from lithium halides can be assumed. %
	}%
	\paragraph{}{%
		Though, upon considering the PLQY and EQE depending on LiTFSI concentration (Fig.\,\ref{PL}\,b) it becomes evident, that another mechanism is present. %
		The different dependence on LiTFSI concentration between EQE and PLQY cannot fully be explained by a surface trap passivation mechanism, since the EQE ($\eta_\mathrm{EQE}$) is proportional to the PLQY: %
		\begin{equation}%
			\eta_\mathrm{EQE}=\eta_\mathrm{out}\cdot q_\mathrm{PLQY}\cdot\gamma.%
		\end{equation}%
		However, at small doping concentrations the drop in EQE is not as pronounced as it is for the PLQY. %
		Other factors only influencing the EQE are the outcoupling factor ($\eta_\mathrm{out}$) and the charge carrier balance ($\gamma$).\cite{Schmidt2017} %
		So an emitter orientation change can enhance the EQE compared to the PLQY, by a limited amount. %
		This change in radiation pattern is observed as shown in Fig.\,\ref{PL}\,d, but the change of orientation is simulated to result in an EQE increase factor of only 1.3. %
		Considering the PeLED fabricated with a 1\wtpct LiTFSI doping, as compared to the one with native \CPB -NCs, the PLQY does not change significantly but the EQE does from about 0.2\,\% to 0.8\,\% by a factor of 4. %
		The orientation alone cannot account for such large enhancement, but the last factor, the charge carrier balance can. %
		Consequently, the doping procedure modifies the charge carrier balance in favour of EQE. %
		Further investigations into that have been made by designing a hole-only device. %
		This device exhibits hole-currents up to a factor 6 higher for the doped sample (see Fig.\,\ref{holeOnly}), implying an increased hole injection of roughly the same factor. %
		The leveraged hole-injection can be ascribed to an improved VB alignment with the adjacent PVK's HOMO level, as measured by UPS and XPS (see Fig\,\ref{UPS}). %
		The injection barrier is reduced from 0.31\,eV to 0.06\,eV, which is in accordance with the increased hole current in the devices. %
		At that, not only the VB onset is shifted but the entire density of states by 0.25\,eV, raising the CB level by the same amount and rendering the \CPB -NC semiconducting thin-film less n-type. % 
		Thus, excess electron injection is reduced as well. %
		The significance of the VB alignment can be seen when comparing the differently coloured LEDs: %
		For the blue CsPbBr\ts2Cl the enhancement of current and EQE is highest, it decreases in green \CPB to no significant current increase for the red MAPbBrI\ts2. %
		The sky-blue CsPbBr\ts2Cl has a VB onset reported to be even lower,\cite{Ravi2016} rendering a potential p-doping even more beneficial as observed. %
		The red device presumably does not benefit by the energy level shift, since because of its small bandgap\cite{Ravi2016} the injection barrier has not been the limiting factor, even though the EQE is still increased significantly. %
	}%
	\paragraph{}{%
		When considering the higher doping concentrations, 5\wtpct and more, a drop in luminance at elevated currents can be observed (see Fig.\,S6). %
		Though the currents are further increasing the luminance does not exceed 20\,cd/m\tS2. %
		A drastic EQE drop occurs at this critical luminance value. %
		This drop is proven to be irreversible by multiple measurement runs of the LEDs. %
		Apparently, there is a degradation induced by an excess of LiTFSI doping. %
		Different interactions of lithium ions with the LHP NCs ranging from residing on an interstitial position over cation exchange (Cs\tS+ for \CPB) to filling lead vacancies have been explored.\cite{Wu2021}
		Upon excess doping multiple of these interactions may mix and change the LHP NCs properties towards less stability and quantum yield.
		Even though this drop in quantum yield is observed in PLQY measurements at 50\wtpct doping only, the drop in operational stability is already dominant for the PeLEDs starting from 5\wtpct. %
		The degradation above a certain luminance together with the fact that \CPB is optically stable even upon strong excitations,\cite{Rain2019,Pierini2020} leads to the conclusion, that the degradation has to originate from an interaction between charge carriers and the excited state. %
		Due to the revealed n-type nature of the LHP NCs and the different current onsets observed in the j-V-L curves, %while considering transport layers' carrier mobilities, 
		it is evident that there is an electron abundance in the device during light-emission.
		Several reports suggest a chemical reduction reaction of lead as a major degradation path in LHPs,\cite{Samu2017,Mulder2021} %
		which may be facilitated by this surplus of electron in PeLEDs during operation. %
		Finally the trion state, consisting of two electrons and one hole, may be the main cause of degradation. %
		LiTFSI seems to have no influence on this proposed electron-induced degradation, since it only plays a minor role in the mitigation of electrons if at all. %
		Its major contribution is ascribed to the easier hole injection into the LHP NCs not necessarily caused by the slightly less n-type energy alignment but rather by the reduced hole injection barrier. %
		As a consequence the degradation seems to affect the EQE in PeLEDs significantly. %
		With the red and blue device the LiTFSI has proven to be very versatile not only increasing the performance of simple all-inorganic tri-bromide LHP NCs, but also in mixed halide or organic perovskite NCs. %
		However the degradation of PeLEDs cannot be prevented by the use of LiTFSI, since its major influence is related to hole-injection and increased PL quantum efficiency.%
	}%
	\section{Conclusion}%
	\paragraph{}{%
        Surface passivation chemistry has already been recognised as key to enhance the performance of LHP NCs regarding its optoelectronic properties. %
        Here we investigate surface treatment with a p-dopant, viz. the organic lithium salt LiTFSI, and show that the efficiency in PeLEDs is tremendously enhanced. %
        We trace this enhancement back to positive effects on three decisive properties considering their EQE: %
        the emitter's quantum yield is proven to be brought close to unity. This is achieved by most surface passivation techniques on LHP NCs. %
        Additionally, the angular emission pattern is changed in favour of better light outcoupling. %
       	While these two effects contribute to the increased efficiency to a certain extent, the most pronounced influence is identified to be the simultaneous tuning of the charge carrier balance to promote hole injection. %
       	Enhancement in these three key factors is a remarkable feature that can be attributed to the LiTFSI treatment. %
        However there are still challenges concerning stability, which is still a major issue for LHP NCs in LEDs in general. %
        At the same time the interaction as well as degradation of LHP NCs in conjuction with LiTFSI is not entirely understood. %
        Nevertheless, a 4-7 fold increase in EQE can be achieved for PeLEDs of various colours and chemistry, rendering the LiTFSI treatment a very promising procedure to further investigate on. %
    }%
	\section{Methods}%
	\subsection{Materials}%
	\paragraph{}{%
		The ITO (Indium Tin Oxide) substrates with dimensions 2 by 2\,cm have been purchased by Kintec (Hongkong) with and without custom pattern of layer thickness 100\,nm on a 23\,nm SiO\textsubscript{2} buffer on a 0.7\,mm thick glass substrate. %
		Fused Silica (SiO\textsubscript{2}) substrates with dimensions 2 by 2\,cm have been bought with a thickness of 0.7\,mm from Nano Quarz Wafer Germany GmbH. %
		PEDOT:PSS is used in a low-conductive ratio of 1:20, having the descriptor CH8000, by Heraeus Germany GmbH \& Co. KG. %
		ZnO has been systhesised from Zincacetate with the sol-gel method.\cite{Hasnidawani2016} %
		\CPB solution (c\,=\,10mg/ml in toluene, ProductID: 900746) and LiTFSI (Lithium bis(trifluormethane)sulfonimide, 99.95\,\% trace metal basis, ProductID: 544094) have been ordered from Merck Germany KGaA. %
		PVK (Poly(9-vinylcarbazole), Mw>10\textsuperscript{5}, ProductID: LT-N4078) has been obtained from Luminescence Technology Corp. (Lumtec, Taiwan).%
	}%
	\subsection{Nanoparticle Preparation}%
	\paragraph{Preparation of oleylammonium halide (OLA-HX)}{%
		To prepare a 1.1\,mmol/ml OLA-HX precursor solution, 10\,ml of oleylamine (OAm) has been placed in a 25-ml three-neck flask and either 1\,ml of concentrated hydrochloric acid (HCl(aq.)) or 1.28\,ml of concentrated hydrobromic acid (HBr(aq.)) has been added slowly. %
		Subsequently, the solidified reaction mixture has been heated at 120\,\textdegree C under nitrogen atmosphere for 2 hours. %
		The reaction temperature has then been increased to 150\,\textdegree C for 30 minutes and afterwards allowed to cool to room temperature. %
		The mixture has been kept in a glovebox and heated to 80\,\textdegree C before injection.%
	}%
	\paragraph{CsPbBr\ts{2}Cl nanocrystals}{%
		CsPbBr\ts2Cl Nanocrystals have been made by a Hot-Injection synthesis using a modified literature method.\cite{Dutta2019} %
		To synthesise 7\,nm CsPbBr\ts2Cl nanocrystals, 49\,mg (0.15\,mmol) Cs\ts2CO\ts3, 67\,mg (0.3\,mmol) PbO and 1.5\,ml oleic acid (OA) have been degassed in 15\,ml ODE in a 50\,ml three-neck flask under reduced pressure at 120\,\textdegree C for 1 hour. %
		The temperature has been increased to 240\,\textdegree C under nitrogen atmosphere, 1\,ml OLA-HBr and 0.5\,ml OLA-HCl precursor have been quickly injected and after one minute the reaction mixture has been cooled to room temperature using an ice-bath (below 180\,\textdegree C). %
		CsPbBr\ts2Cl NCs have been collected by centrifuging the suspension (7000\,rpm, 10\,min.), decanting the supernatant, and collecting the precipitate. %
		The precipitate has been centrifuged again without addition of a solvent (7000\,rpm, 5\,min.), and the resulting supernatant has been removed with a syringe, to separate the traces of residual supernatant. %
		The precipitate has been dissolved in 2\,ml hexane and centrifuged again (2500\,rpm, 5\,min.) to remove aggregates and larger particles. %
		The resulting supernatant has been filtered through a 0.2\,\textmu m PTFE syringe filter and stored as stock solution inside of a glovebox with a typical concentration of 25\,mM following Maes et al..\cite{Maes2018} %
	}%
	\paragraph{LiTFSI treatement}{%
		The LiTFSI solutions are diluted from a stock which is created by dissolving 200\,mg LiTFSI with 2\,ml dimethylformamide (DMF) and 18\,ml chlorobenzene (CB) by stirring overnight resulting in a volume concentration of 10\,mg/ml. %
		Dilution to 1, 0.1 and 0.01\.mg/ml concentration is done with CB only.\\
		Equal volumes of LHP NC solution and LiTFSI solutions are mixed to obtain a LiTFSI-doped solution. %
		Mixing equal volumes of 10\,mg/ml LHP NC and 10\,mg/ml LiTFSI yields a 5\,mg/ml LHP NC solution with 50\,wt.\,\% LiTFSI doping. %
		Analogously, a 9.09, 0.990 and 0.099\,wt.\,\% doped solution is obtained by using 1, 0.1 and 0.01\,mg/ml concentrated LiTFSI solution and mixing with the 10mg/ml LHP NC solution by equal volumes. %
		All LHP NC solutions are created within a nitrogen filled glovebox and have exhibited stable luminescence for at least 3 months at room temperature.
	}%
	\subsection{Sample Preparation}%
	\paragraph{Photoluminescence}{%
		Fused Silica is used as a substrate for all PL measurements, that is PLQY, TRPL and ADPL. %
		The substrates are spin-coated on in a nitrogen-filled glovebox by dropping 50\,\textmu l (p.r.n. LiTFSI-doped) LHP NC solution before starting the rotation of the spin-coater. %
		After a settling time of 30 seconds it is accelerated to 500\,rpm and kept at that speed for another 30 seconds. %
		To remove residuals from the edges, spinning for 5 seconds at 2000\,rpm is applied before stopping the procedure.%
	}%
	\paragraph{UPS/XPS, SEM, LEDs and hole-only devices}{%
		For UPS/XPS and SEM the unpatterned, and for the electrical devices the patterned ITO substrates are used. %
		Initally, PEDOT:PSS is spin-coated in the cleanroom, at 4000\,rpm for 30s, and heated on a hotplate at 130\,\textdegree C for 15 minutes, resulting in a smooth approx. 40\,nm thin film. %
		The samples are transferred to a nitrogen-filled glovebox immediately. %
		As second layer, the approx. 20\,nm thin PVK film is deposited by spin-coating a 3\,mg/ml concentrated PVK-chlorobenzene solution at 3000\,rpm for 30 seconds and heating at 175\,\textdegree C for 30 minutes. %
		The LHP NC solutions are spin-coated as described for the PL samples, after cooling the substrate to room-temperature, resulting in a closed film of about 2-3 monolayers (effective thickness approx. 20\,nm).\\
		The sample in current state, that is the bottom-half LED, is used for UPS/XPS and SEM investigations. %
		For devices, the samples are transferred without ambient exposure into a high vacuum chamber, with a pressure smaller than 10\textsuperscript{-6}\,mbar. %
		For hole-only, 10\,nm HATCN (rate: 50\,pm/s), for LEDs 55\,nm B3PYMPM (rate: 100\,pm/s) followed by 0.5\,nm LiF (rate: 10\,pm/s) is evaporated. %
		The devices are finished with a 60\,nm (rate: 100\,pm/s) thick Aluminium cathode, also deposited by thermal deposition.%
	}%
	\subsection{Measurement details}%
	\paragraph{PLQY}{%
		The PLQY is determined by a two-step-method measurement featuring a BaSO\ts4 coated integrating sphere.\cite{Leyre2014} %
		The excitation source is a HeCd laser's 442\,nm light. %
		The excitation signal as well as the samples fluorescence is collected with a fiber and guided into a Princeton Instruments Acton2300i spectrometer, which is connected to a nitrogen cooled CCD camera, that is Princeton Instrument's Pylon BRX100. %
		An absolute calibration of the integrating-sphere-CCD system has been performed with a lamp calibrated for spectral irradiance according to the NIST standard by GigaHertz Optik GmbH Germany. %
		With that system spectra could evaluated to its amount of photons and consequently a PLQY is calculated.%
	}%
	\paragraph{ADPL}{%
		The ADPL measurement and analysis were performed as previously reported by our group.\cite{Morgenstern2020} The spectrum is recorded with the same CCD-Spectrometer system as explained for the PLQY.%
	}%
	\paragraph{TRPL}{%
		Transient Photoluminescence was recorded by the C5680 streak camera system by Hamamatsu, after being delayed by DG535 by Stanford Instruments while being spectrally analysed by a Acton Spectra Pro 2300i. %
		Excitation was done with the EKSPLA PT400 Diode Pumped Solid State laser set to wavelength 355\,nm.%
	}%
	\paragraph{UPS/XPS}{%
		Ultraviolet photoelectron spectroscopy (UPS) measurements have been conducted using a SPECS PHOIBOS 100 hemispherical electron analyser equipped with a monochromatised helium discharge lamp (21.22\,eV). The UV flux has been attenuated significantly by the monochromator to avoid UV-induced sample degradation. A sample bias of -10\,V has been applied to acquire the secondary electron cutoff spectra. The base pressure of the analysis chamber has been kept below 10\tS{-9}\,mbar. X-ray photoelectron spectroscopy (XPS) measurements have been performed at a JEOL JPS-9030 ultrahigh vacuum system (base pressure of 10\tS{-9}\,mbar) using monochromatised Al K\ts{$\alpha$} (1486.6\,eV) radiation. Anode power of 30\,W was applied for XPS measurements, which has not been found to induce noticeable sample degradation.%
	}%
	\paragraph{LEDs and hole-only devices}{%
		j-V(-L) curves are recorded with a Keithley 2612B Source Meter Unit (SMU). %
		A Photodiode of known diameter at known distance is used for Luminance detection. %
		The electroluminescent spectrum is taken with the Phelos system by fluxim AG (Switzerland). %
		With that and a lambertian approximation the EQE is determined. %
		The integrity of the lambertian approximation has been ensured by random sampling with a calibrated integrating sphere (same setup as for PLQY). %
		The sampling revealed that the photodiode's and consequently in this manuscript reported EQE is about 10\,\% underestimated, relatively. %
		For instance the EQE of the 9\,wt.\,\% LiTFSI-doped sample in Fig.\,\ref{PL}\,b has shown 1.1\,\% EQE in the integrating sphere while it has been 1\,\% in the photodiode-setup.%
	}%
	\begin{acknowledgement}%
		This work was funded by Deutsche Forschungsgemeinschaft (DFG) within Priority Program SPP2196 ("Perovskite Semiconductors: From Fundamental Properties to Devices") under project nos. 424708673 and 423749265, as well as by their Heisenberg Program under grant SCHE1905/9-1. M.M. acknowledges postdoctoral support from the Alexander von Humboldt Foundation.%
	\end{acknowledgement}%
	\begin{suppinfo}%
		The following supporting material is available:%
		\begin{enumerate}%
			\item TRPL-fits and details (Fig.\,S1\,\&\,S2).%
			\item ADPL-fits and details (Fig.\,S3).%
			\item SEM images (Fig.\,S4).%
			\item XPS spectra (Fig.\,S5).%
			\item Details on DFT calculations with respective structures for the slab models.%
			\item \CPB NC LEDs at all LiTFSI concentrations (Fig.\,S6) and MAPbBrI\ts2 NC LED details (Fig.\,S7).%
            \item Additional details on chemicals and preparation.%
		\end{enumerate}%
	\end{suppinfo}%
	\bibliography{2022_Naujoks_QE_Enhancement_of_LHP_LEDs_by_LiTFSI_acsami.bib}%
\end{document}

\maketitle%
    \section{Transient Photoluminescence Spectroscopy:}%
	\begin{figure}[hbt]%
		\begin{minipage}{0.5\textwidth}%
			\begin{overpic}[width=\linewidth]{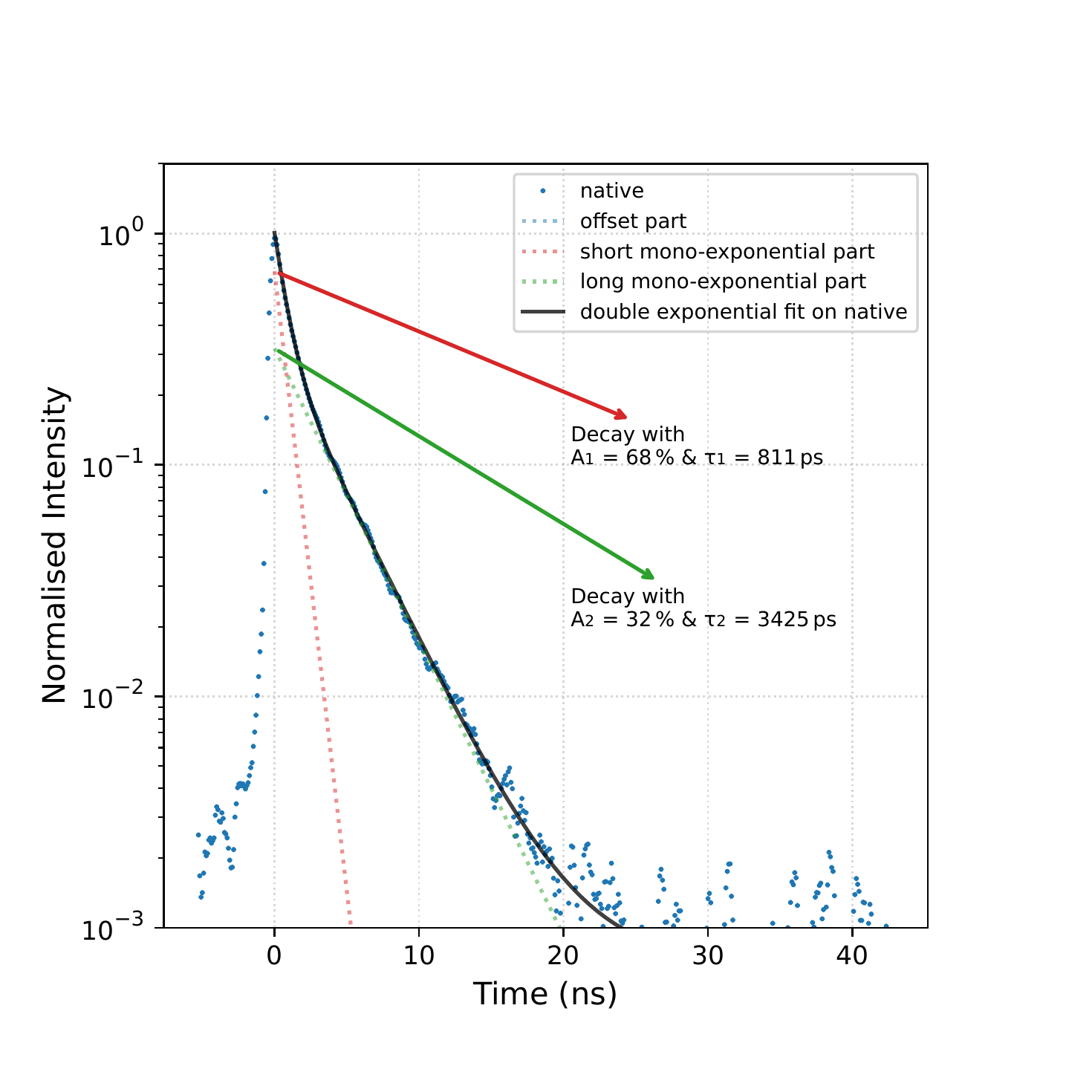}%
				\put (5,95) {\scriptsize a)}%
			\end{overpic}%
		\end{minipage}{}%
		\begin{minipage}{0.5\textwidth}%
			\begin{overpic}[width=\linewidth]{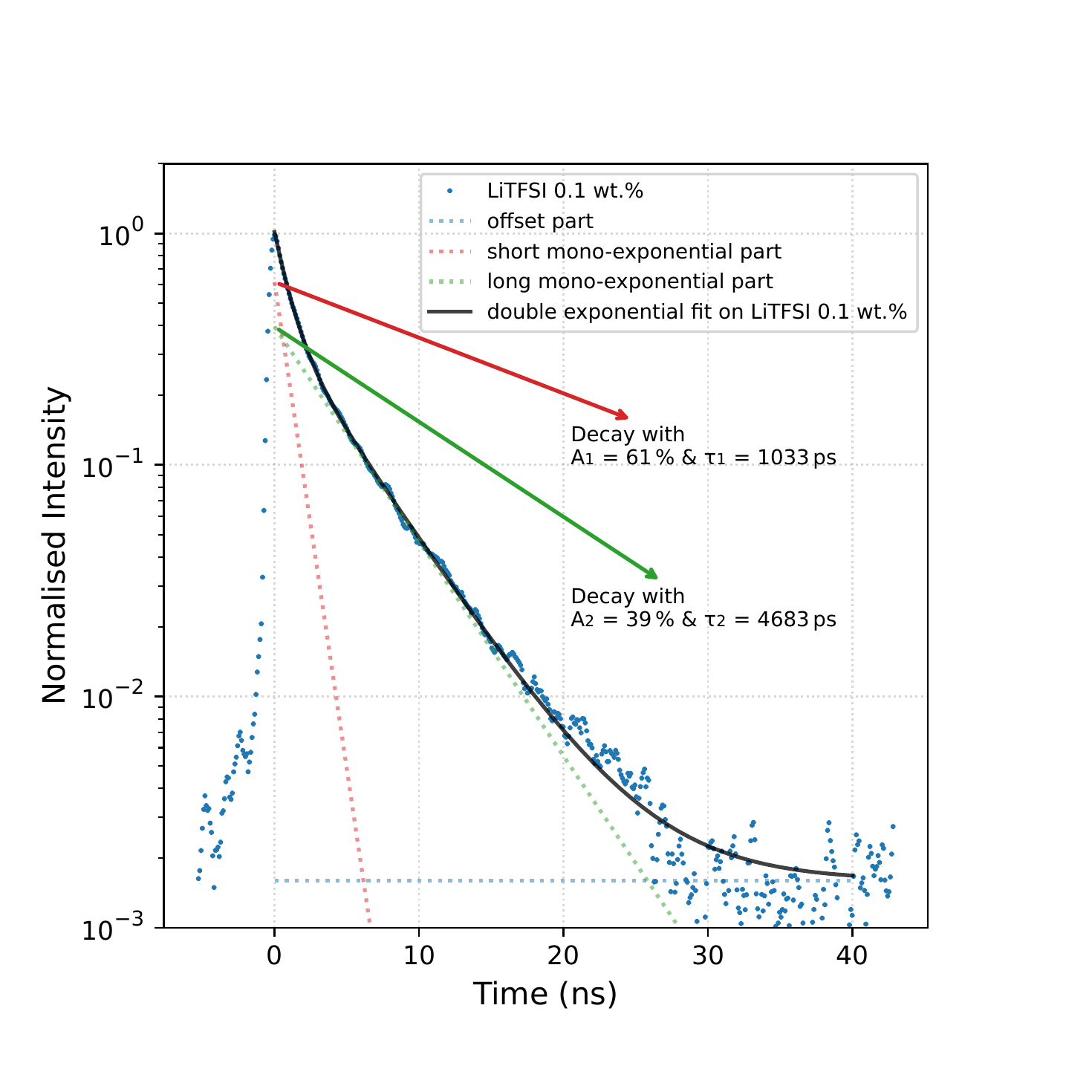}%
				\put (5,95) {\scriptsize b)}%
			\end{overpic}%
		\end{minipage}\\%
		\begin{minipage}{0.5\textwidth}%
			\begin{overpic}[width=\linewidth]{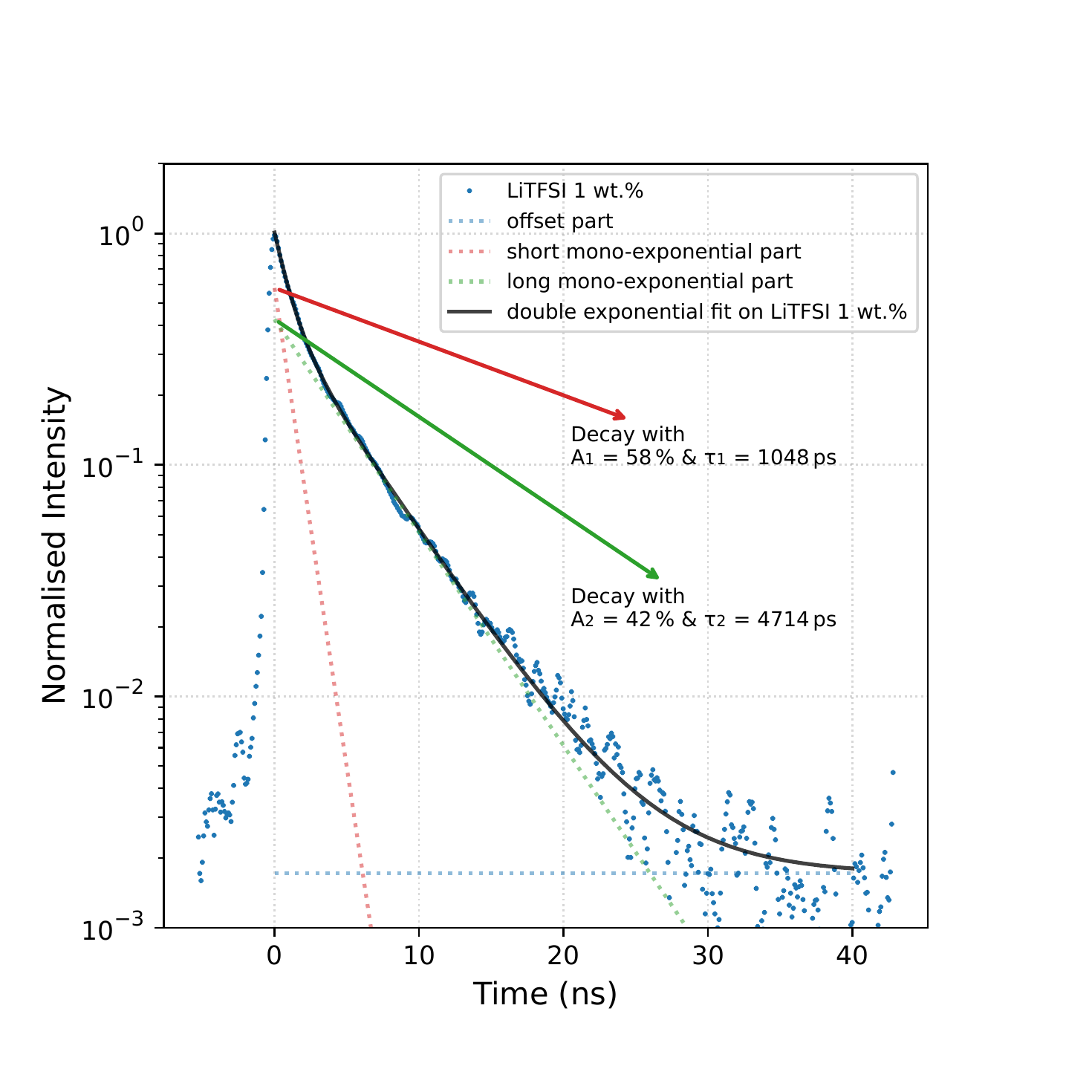}%
				\put (5,95) {\scriptsize c)}%
			\end{overpic}%
		\end{minipage}{}%
		\begin{minipage}{0.5\textwidth}%
			\begin{overpic}[width=\linewidth]{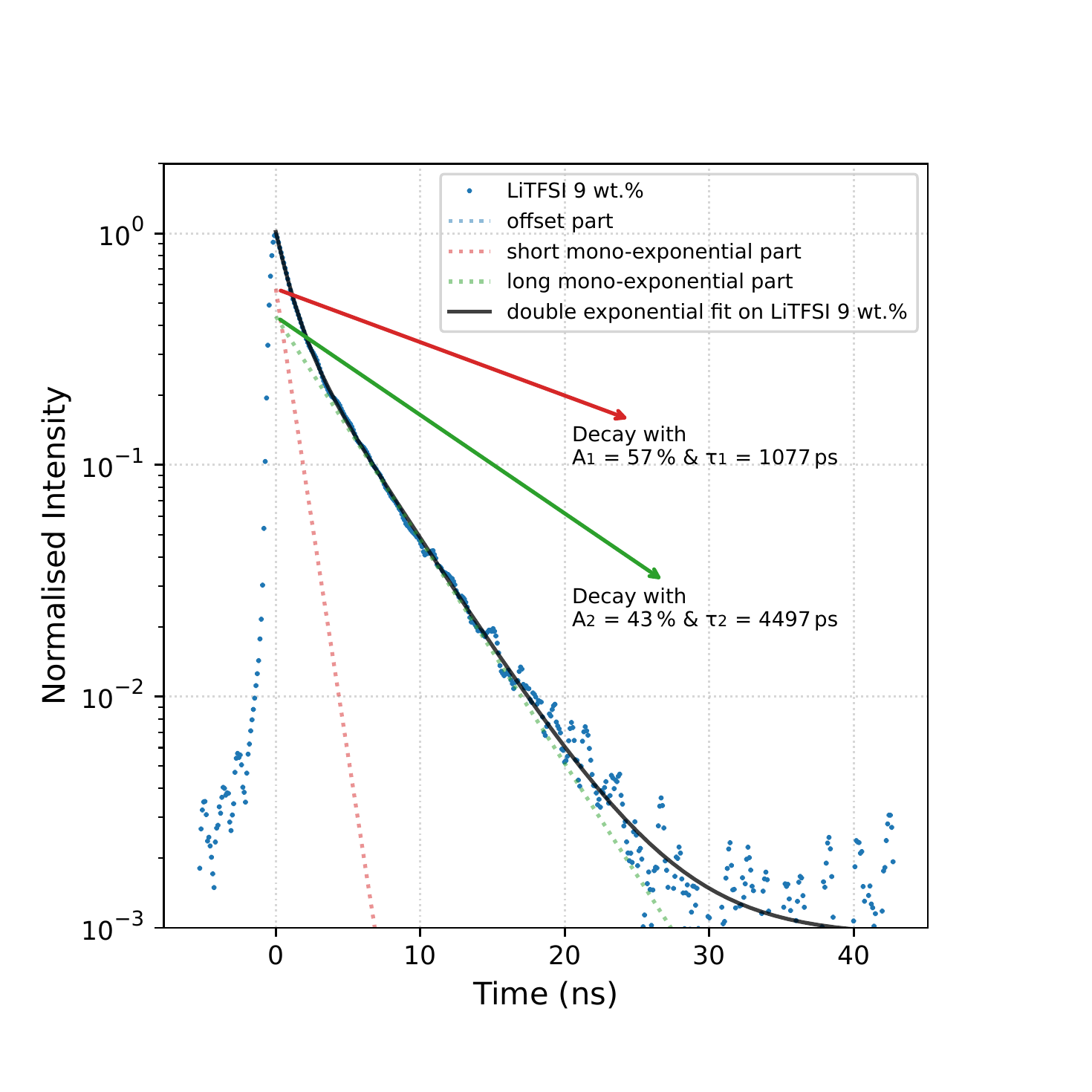}%
				\put (5,95) {\scriptsize d)}%
			\end{overpic}%
		\end{minipage}%
		\caption{Transient PL of a \CPB -NC film with a) no, b) 0.1\,wt.\,\%, c) 1\,wt.\,\% and d) 9\,wt.\,\% LiTFSI doping}%
		\label{TRPL0-10}%
	\end{figure}%
	\begin{figure}[hbt]%
		\begin{minipage}{0.5\textwidth}%
			\begin{overpic}[width=\linewidth]{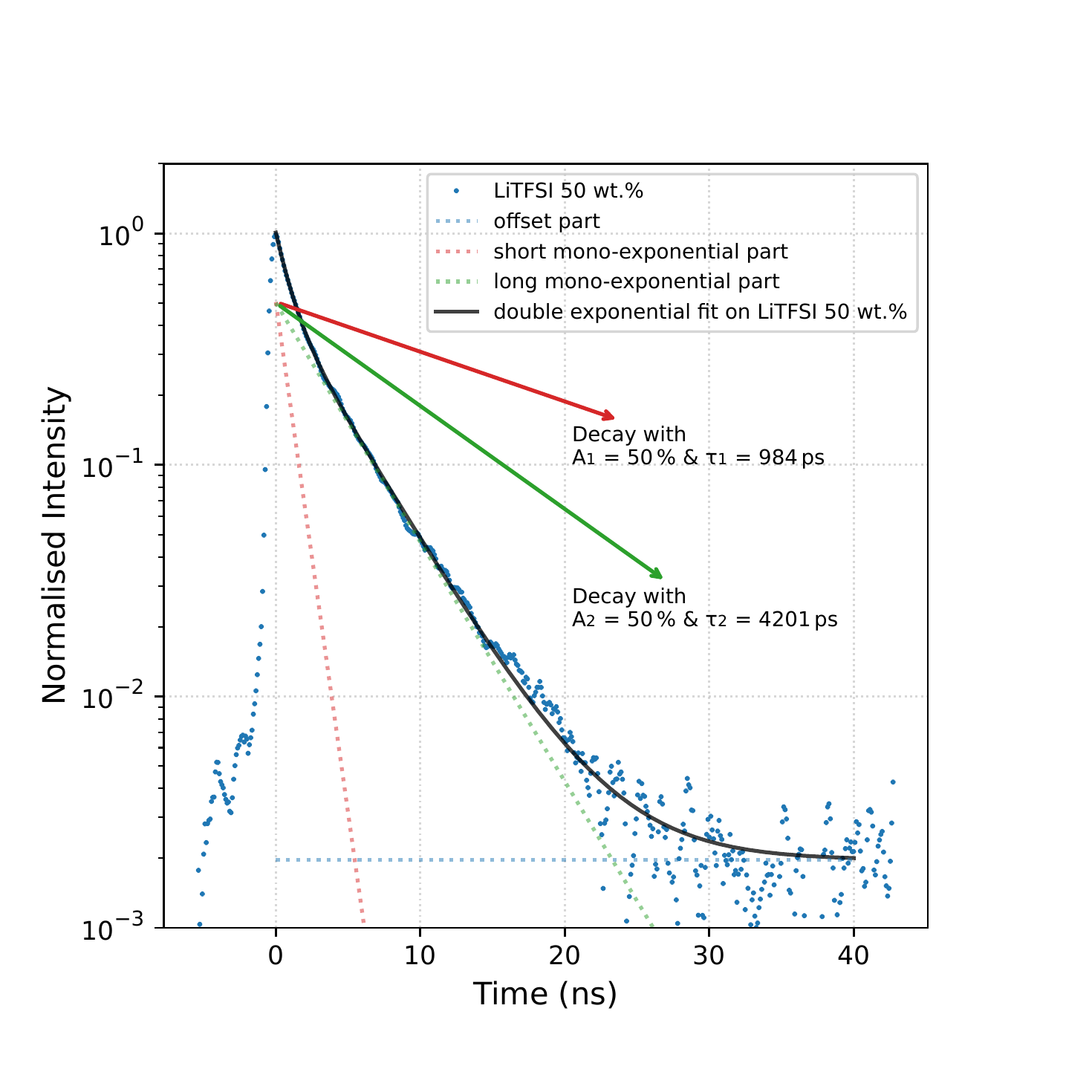}%
				\put (5,95) {\scriptsize a)}%
			\end{overpic}%
		\end{minipage}{}%
		\begin{minipage}{0.5\textwidth}%
			\begin{overpic}[width=\linewidth]{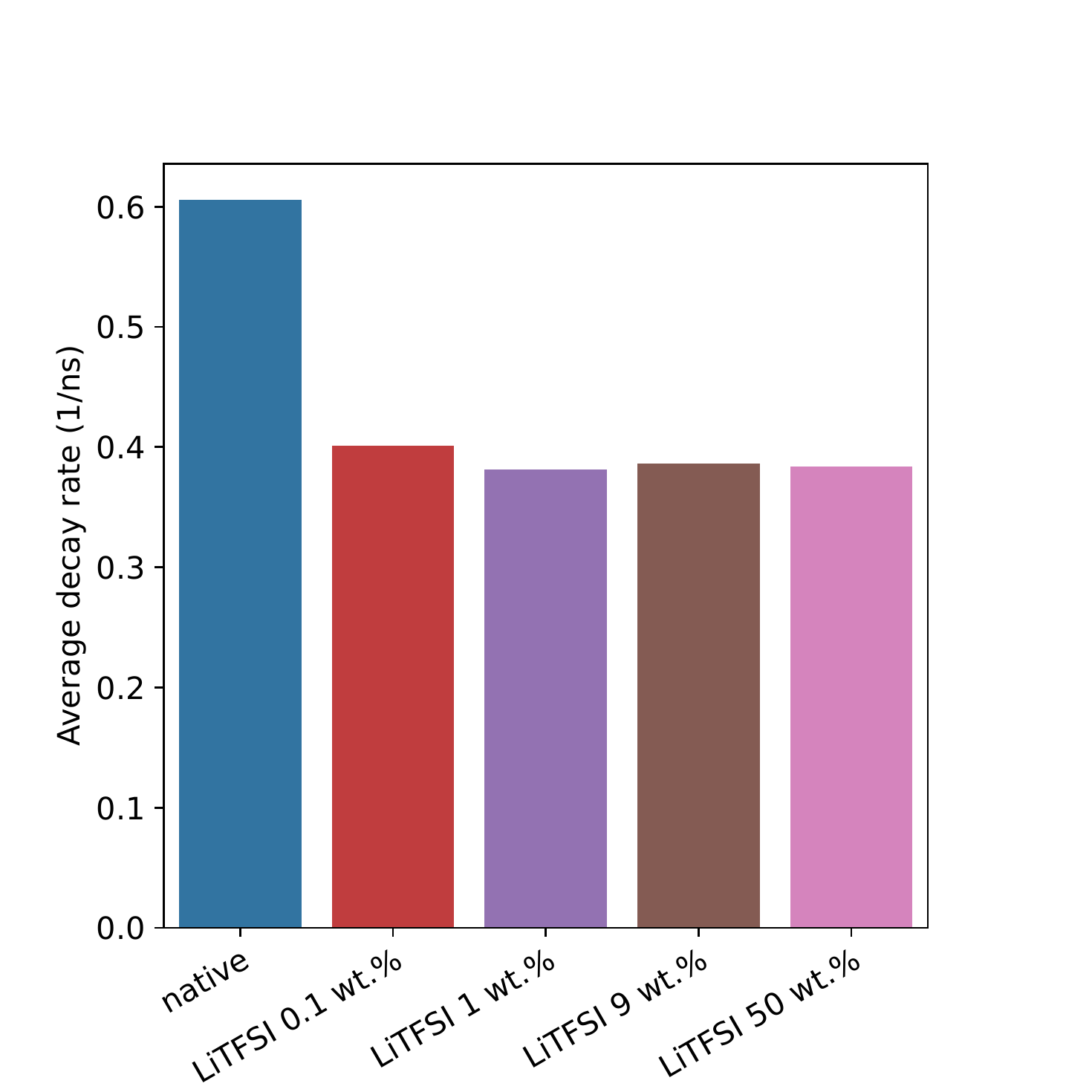}%
				\put (5,95) {\scriptsize b)}%
			\end{overpic}%
		\end{minipage}\\%
		\begin{minipage}{0.5\textwidth}%
			\begin{overpic}[width=\linewidth]{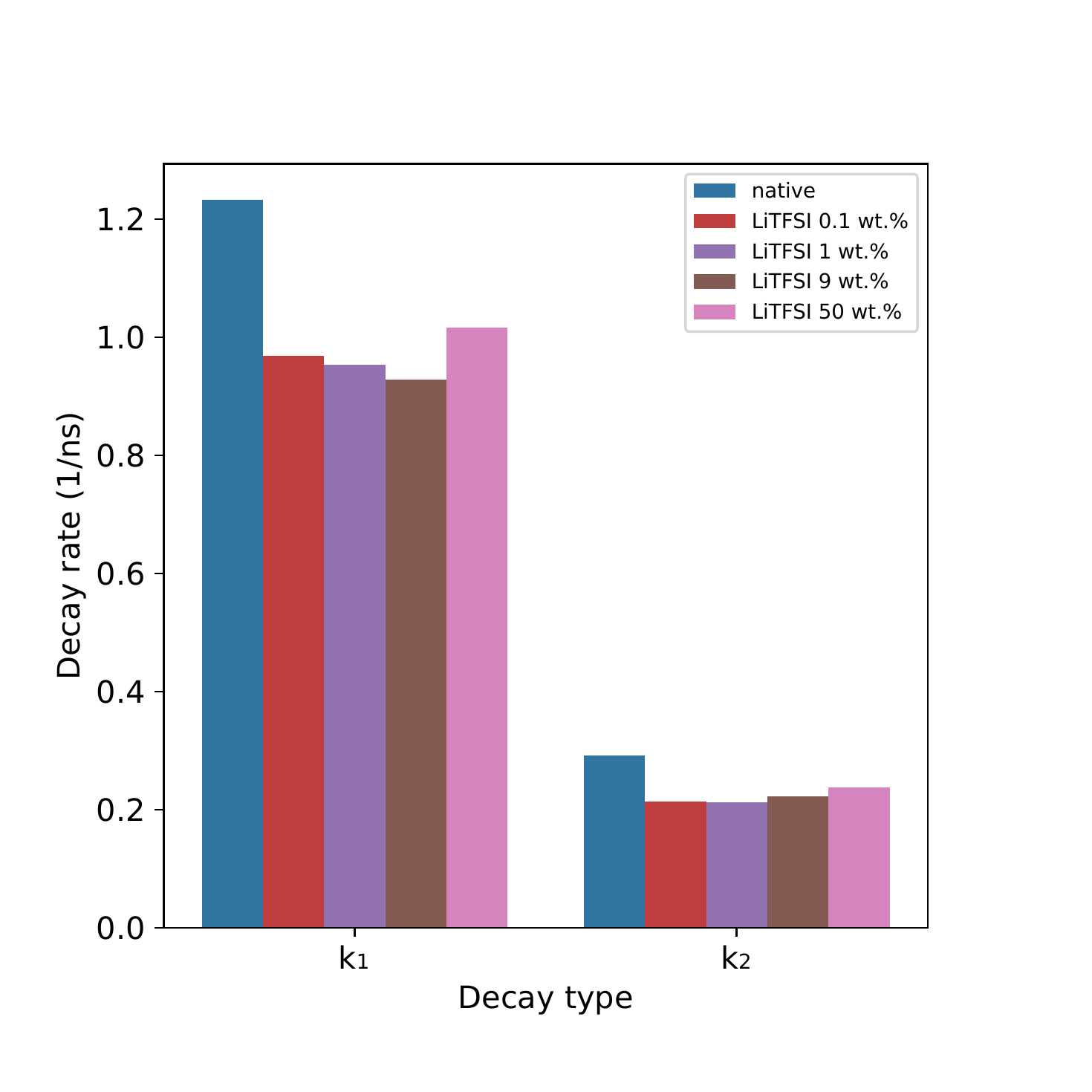}%
				\put (5,95) {\scriptsize c)}%
			\end{overpic}%
		\end{minipage}{}%
		\begin{minipage}{0.5\textwidth}%
			\begin{overpic}[width=\linewidth]{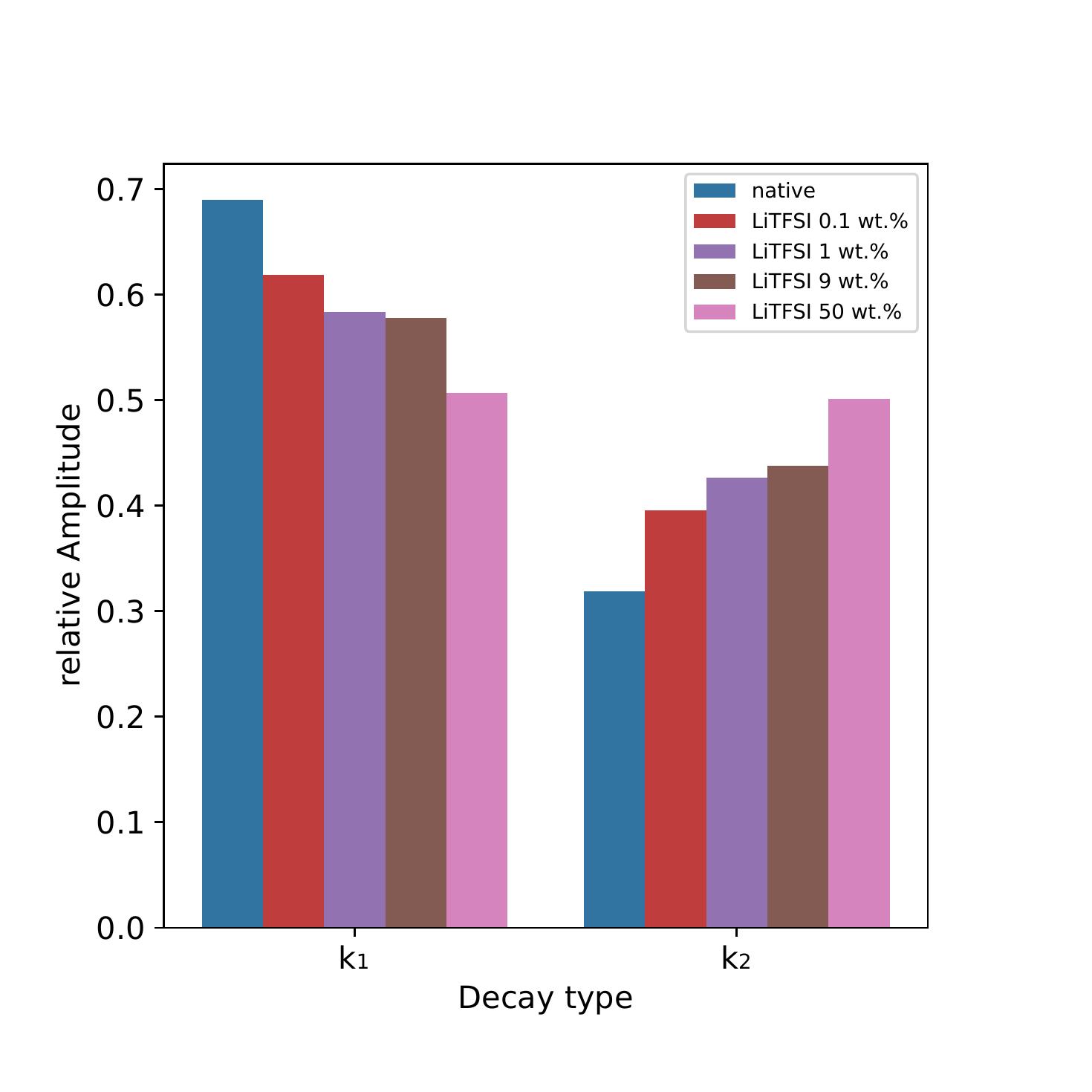}%
				\put (5,95) {\scriptsize d)}%
			\end{overpic}%
		\end{minipage}%
		\caption{Transient PL of a \CPB NC film with a) 50\,wt.\,\% LiTFSI doping. b) average decay rate, c) individual decay rates and d) relative amplitudes against LiTFSI doping concentration.}%
		\label{TRPL50-bars}%
	\end{figure}%
	\paragraph{}{%
	    The normalised transient PL data has been fitted with a double exponential decay function%
	    \begin{equation}%
	    	\mathrm{I(t)}=\mathrm{A}_1\cdot\exp\left[-\frac{t}{\tau_1}\right]+\mathrm{A}_2\cdot\exp\left[-\frac{t}{\tau_2}\right],%
	    \end{equation}%
	    with individual amplitudes $\mathrm{A}_\mathrm{x}$ and lifetimes $\tau_\mathrm{x}$. The reciprocal lifetimes yield the respective decay rates:%
	    \begin{equation}%
	    	\mathrm{k}_\mathrm{x}=\frac{1}{\tau_\mathrm{x}}.%
	    \end{equation}%
	    	Additionally, for comparison, an amplitude-weighted average lifetime $\tau_\mathrm{avg}$ is defined (with the amplitudes summed being one):%
	    \begin{equation}%
	    	\tau_\mathrm{avg}=\sum_{i=0}^{n}\mathrm{A}_i\cdot\tau_i.%
	    \end{equation}%
	    The data shows no clear trend within the concentration sweep apart from the significant jump from undoped to the doped ones. %
	    Only the amplitude, that is, the dominance of the slower decay increases with LiTFSI concentration.%
	}%
	\clearpage%
    \section{Angular Dependent Photoluminesence Spectroscopy:}%
\begin{figure}[bth]%
        \begin{minipage}{0.5\textwidth}%
            \begin{overpic}[width=\linewidth]{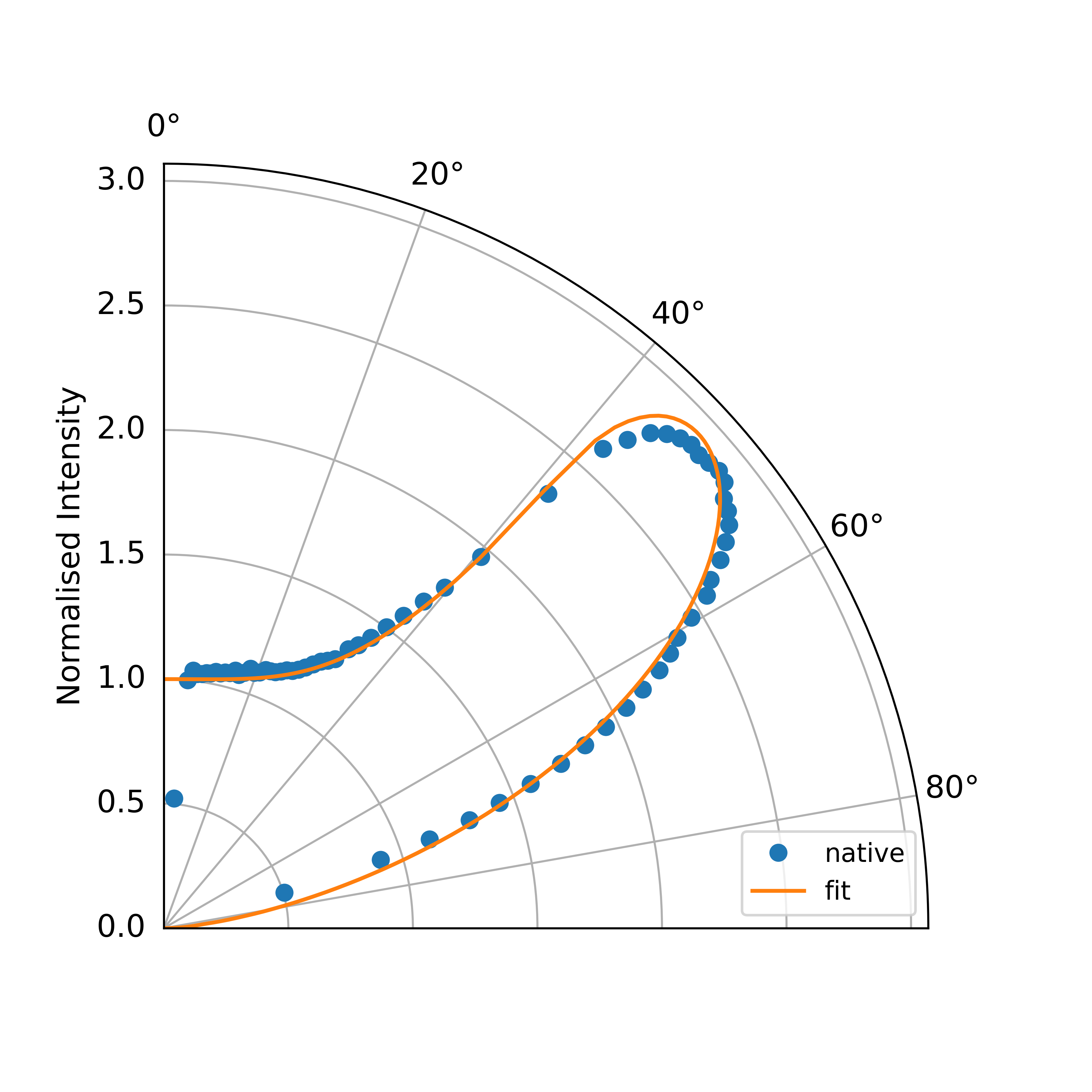}%
                \put (5,95) {\scriptsize a)}%
            \end{overpic}%
        \end{minipage}{}%
        \begin{minipage}{0.5\textwidth}%
            \begin{overpic}[width=\linewidth]{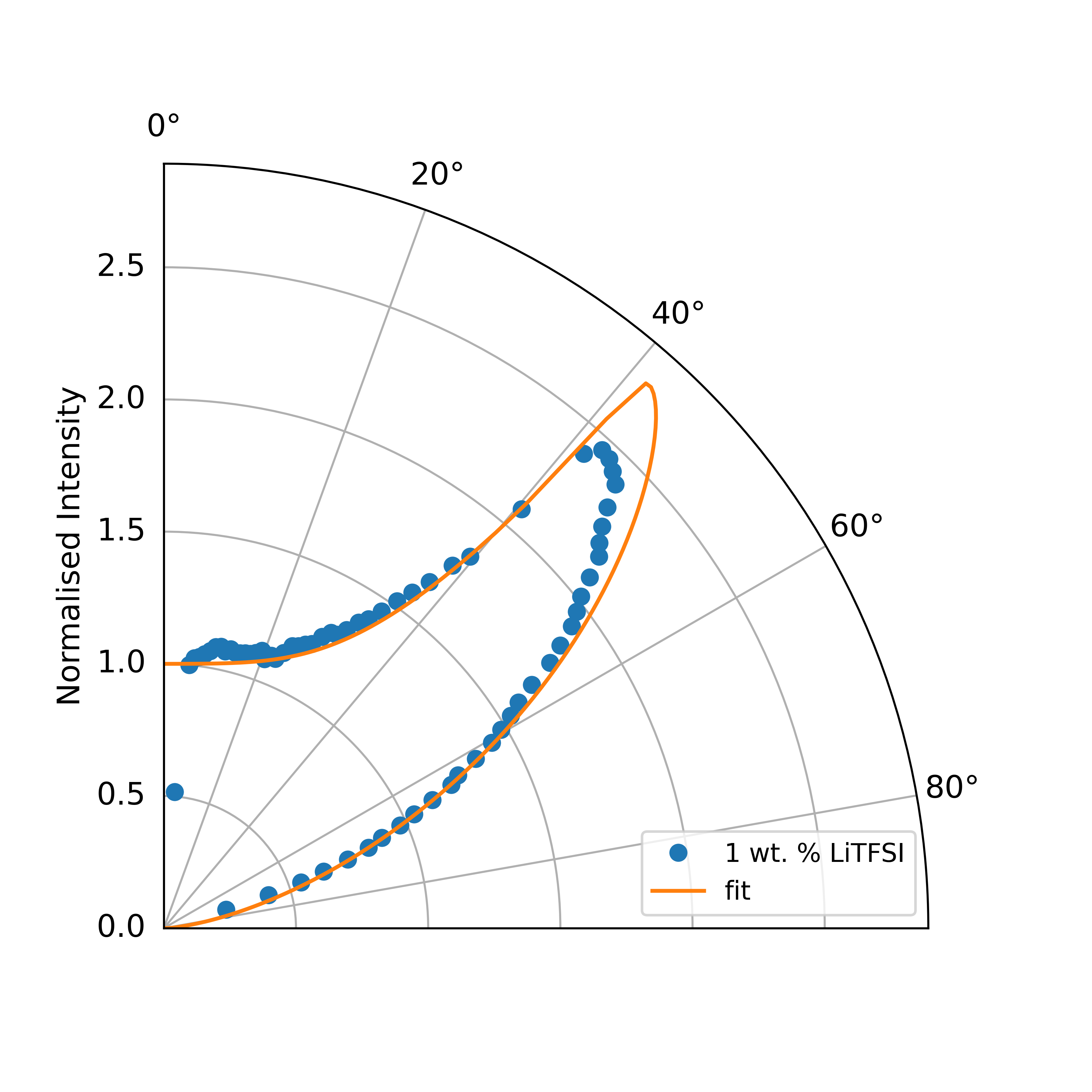}%
                \put (5,95) {\scriptsize b)}%
            \end{overpic}%
        \end{minipage}\\%
        \begin{minipage}{0.5\textwidth}%
            \begin{overpic}[width=\linewidth]{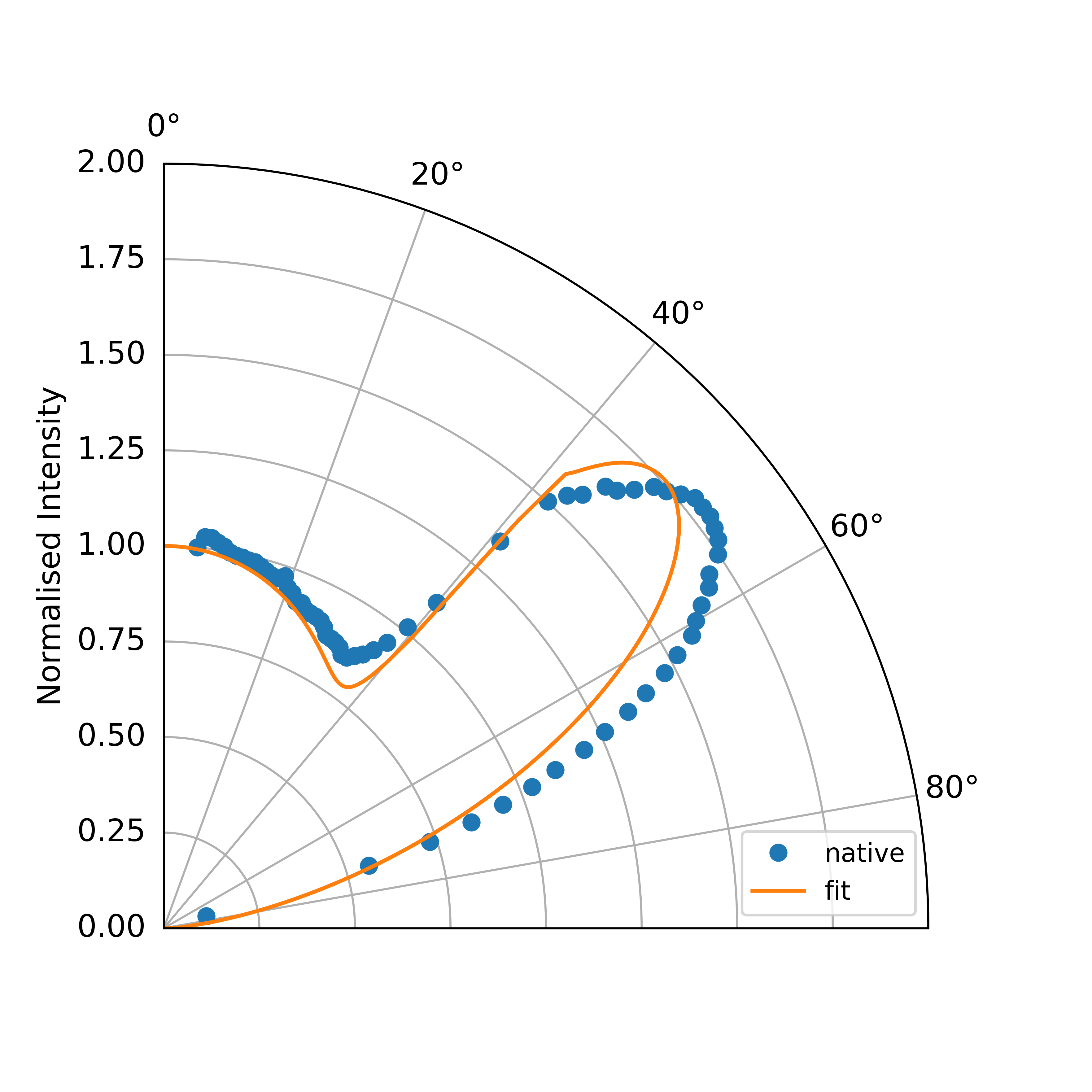}%
                \put (5,95) {\scriptsize c)}%
            \end{overpic}%
        \end{minipage}{}%
        \begin{minipage}{0.5\textwidth}%
            \begin{overpic}[width=\linewidth]{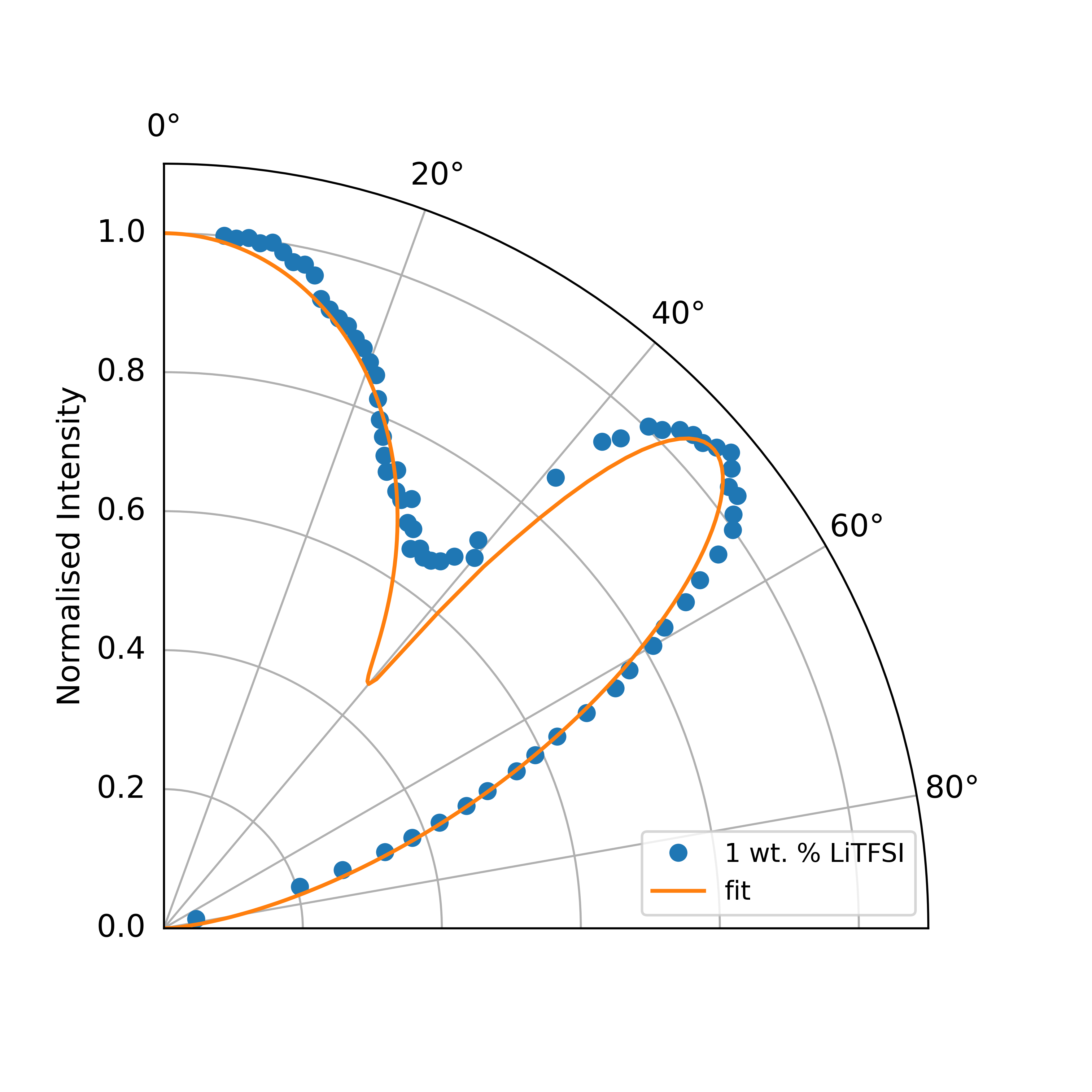}%
                \put (5,95) {\scriptsize d)}%
            \end{overpic}%
        \end{minipage}%
        \caption{s-polarised ADPL spectrum with a matrix-transfer fit of a) native b) 1\,wt.\,\% LiTFSI doped \CPB thin-film on glass (bk7); the corresponding curves for p-polarised ADPL are shown in c) and d).}%
        \label{ADPL}%
    \end{figure}

\begin{figure}[bth]%
		\begin{minipage}{0.5\textwidth}%
			\centering%
			\begin{overpic}[width=0.9\linewidth]{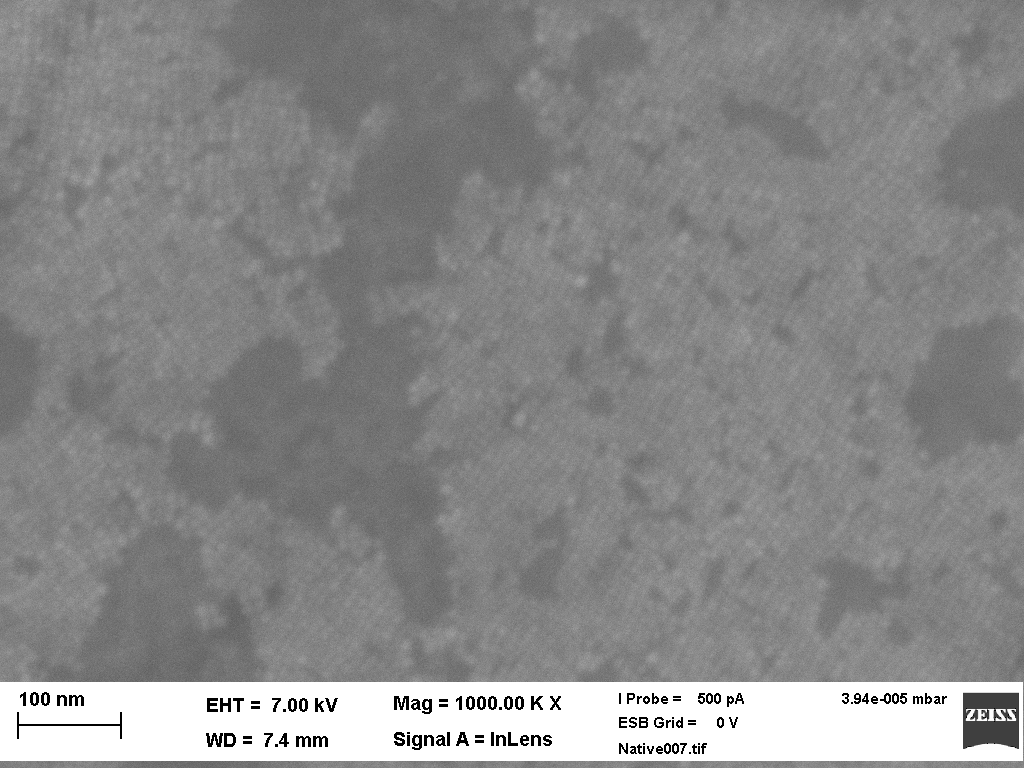}%
				\put (-5,80) {\scriptsize a)}%
			\end{overpic}%
		\end{minipage}{}%
		\begin{minipage}{0.5\textwidth}%
			\centering%
			\begin{overpic}[width=0.9\linewidth]{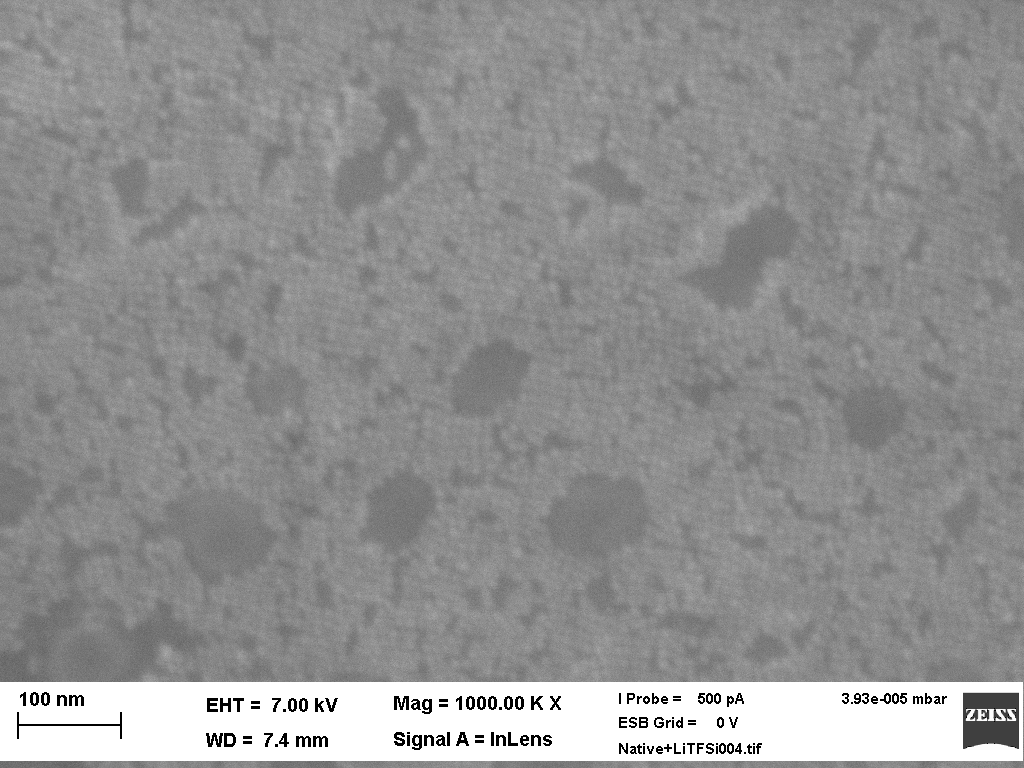}%
				\put (-5,80) {\scriptsize b)}%
			\end{overpic}%
		\end{minipage}%
		\caption{%
			SEM image of \CPB NCs spin-coated on PVK a) without and b) with 1\,wt.\,\% LiTFSI doping. %
			The dark areas have been verified to be another layer of LHP NCs by EDX. %
			Nevertheless the LiTFSI doped sample shows slightly better yet comparable film coverage.%
		}%
		\label{SEM}%
	\end{figure}

    \paragraph{}{%
	    Angular Dependent Photoluminesence Spectroscopy (ADPL) data is recorded in two steps: the orthogonal (s-pol.) and the parallel (p-pol.) part of the PL signal is recorded sequentially. %
	    The p-pol. part contains contributions from transition dipole moments (TDM) that are perpendicular (p\ts{$\perp$}) as well as parallel (p\ts{$\|$}) to the substrate surface, and thus provides the relevant information about the TDM orientation. 
	    Therefore it is used for determining the orientation parameter according to equation:%
	    \begin{equation}%
	        \theta=\frac{\mathrm{p}_\perp}{\sqrt{\mathrm{p}_\perp^2+\mathrm{p}_\|^2}}.%
	    \end{equation}%
		$\theta$ can take values between 0 and 1, while the former denotes fully horizontal orientation and the latter fully vertical orientation. %
		Isotropic orientation is obtained if one third of the TDMs is perpendicular, that is $\theta=\frac{1}{3}$, see Fig.\,\ref{ADPL}\,c\,\&\,d. %
		More details on this method can be found in the publication of Jurow et al.\cite{Jurow2017}, as well as in the original publication for organic semiconductors by Frischeisen et al.\cite{Frischeisen2011}\\%
		The s-pol. part, however, contains only contributions of the TDMs that are parallel to the substrate; %
		it is thus only used for fitting the layer thickness and verifying the refractive index of the LHP NC layers, %
		Fig.\,\ref{ADPL}\,a\,\&\,b: d\ts{native}$\,=\,31\,$nm and n\ts{native}$(\lambda=510\,\mathrm{nm})=1.75$, %
		respectively d\ts{LiTFSI}$\,=\,10\,$nm and n\ts{LiTFSI}$(\lambda=510\,\mathrm{nm})=1.75$. %
		The thickness difference can be explained by a deviating film coverage on fused silica compared to on PVK (as shown in Fig.\,\ref{SEM}).%
	}%
	\clearpage%
	\section{Scanning electron microscopy (SEM):}%
	\section{XPS:}%
	\begin{figure}[bth]%
	    \begin{minipage}{0.5\textwidth}%
	        \begin{overpic}[width=\linewidth]{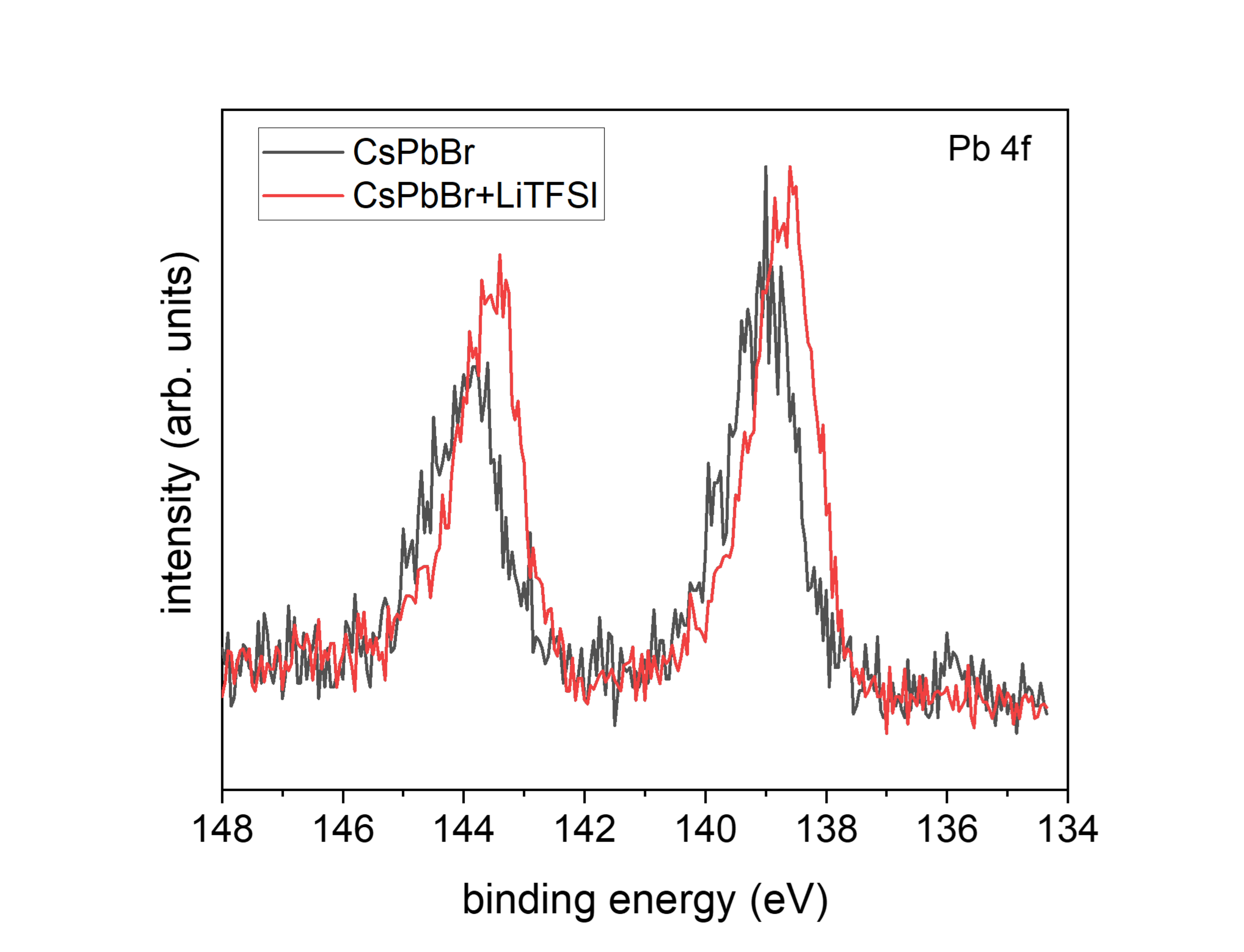}%
	            \put (5,75) {\scriptsize a)}%
	        \end{overpic}%
	    \end{minipage}{}%
	    \begin{minipage}{0.5\textwidth}%
	        \begin{overpic}[width=\linewidth]{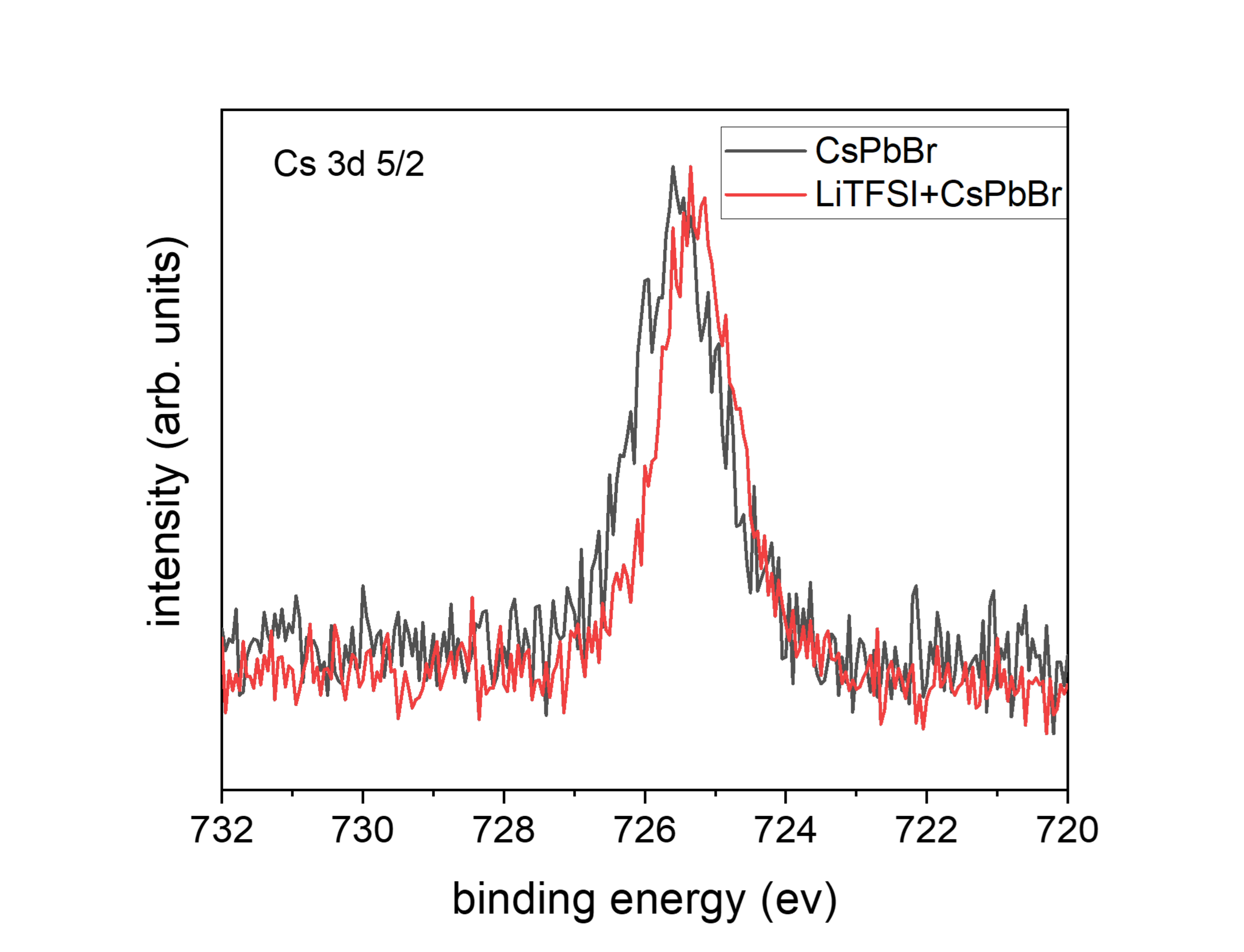}%
	            \put (5,75) {\scriptsize b)}%
	        \end{overpic}%
	    \end{minipage}%
	    \caption{XPS spectra considering the core orbitals of \CPB at a) Pb's 4f and b) Cs 3d 5/2. Even the core levels are shifted by the same amount as the VBMs shown in the main publication of around 0.25\,eV}%
	    \label{XPS}%
	\end{figure}%
	\clearpage%
	\section{Density functional calculations}%
	\paragraph{}{%
		Slab models are constructed from orthorhombic \CPB structure considering a 1×2×3 supercell having 5 Pb–Br layers and exposing the Cs–Br terminated surface. %
		Consecutive slabs in the c-direction have been separated by a vacuum of approximately 20\,\AA  to ensure decoupling with its periodic image. %
		Periodic calculations are performed using a plane wave basis set implementation of density functional theory within the Vienna Ab initio Simulation Package (VASP, version 6.1)\cite{Kresse1996} employing the PBE exchange-correlation functional,\cite{Perdew1996} and van der Waals interactions have been incorporated employing Grimme’s D3 method.\cite{Grimme2010} %
		The valence-core interactions are described with the projected augmented wave (PAW) method.\cite{Kresse1996a, Kresse1999} %
		Valence electrons of each type of atom considered during calculations are: N (5), S (6), O (6) C (4), F (7), Li (3), Cs (9), Pb (14), and Br (7). %
		A plane-wave energy cutoff of 520\,eV is used in all calculations. %
		Forces of each atom smaller than 0.02\,eV/Å are used during geometry relaxation. %
		The structural relaxation has been done by sampling the Brillouin zone over a 3×3×1 k-point grid centered at the $\Gamma$-point. %
		Structure visualisation and the projected density of states plots are performed using the VESTA\cite{Momma2011} and Sumo\cite{Ganose2018} packages. %
		VASPKIT\cite{Wang2021t} is used for plotting planar average electrostatic potentials and obtain vacuum level energy.%
	}%
    \begin{table}
        \begin{tabular}{c|ccc}
            & Vacuum-Level (eV)&E\ts{Fermi} (eV)&Workfunction (eV)\\
            \hline
            model A&2.87&-1.64&4.51\\
            model B&3.71&-1.83&5.54\\
            model C&4.23&-1.97&6.20\\
            model D&3.83&-1.87&5.69
        \end{tabular}
        \caption{Calculated Energy levels for the models elaborated in the main article.}%
        \label{energytable}%
    \end{table}
	\clearpage%
	\section{PeLEDs:}%
	\begin{figure}[bth]%
	    \begin{minipage}{0.5\textwidth}%
	        \begin{overpic}[width=\linewidth]{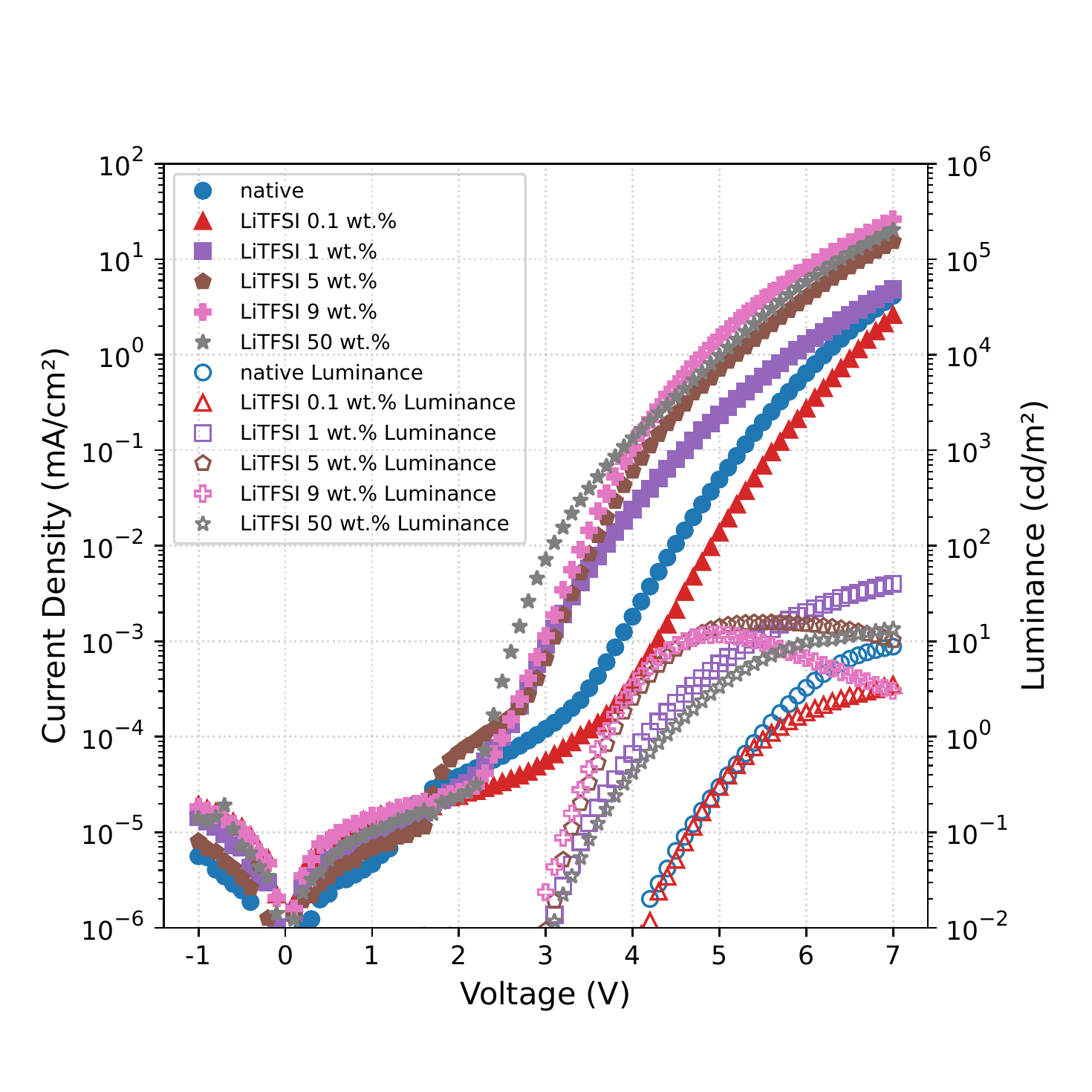}%
	            \put (5,95) {\scriptsize a)}%
	        \end{overpic}%
	    \end{minipage}{}%
	    \begin{minipage}{0.5\textwidth}%
	        \begin{overpic}[width=\linewidth]{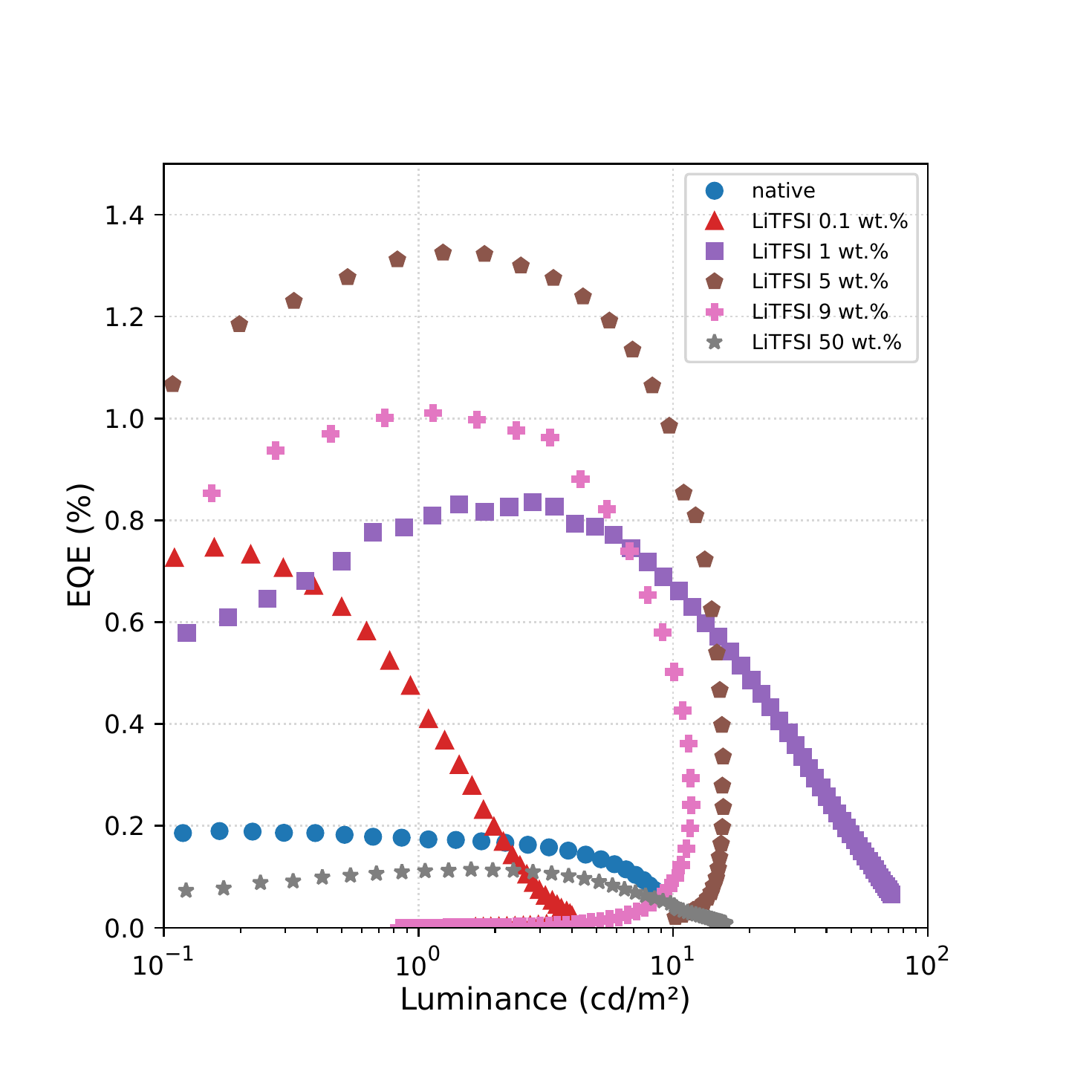}%
	            \put (5,95) {\scriptsize b)}%
	        \end{overpic}%
	    \end{minipage}%
	    \caption{a): j-V-L curves of \CPB NC-LEDs with and without various LiTFSI dopings, with b) its corresponding EQE vs current density curves.}%
	    \label{gLED}%
	\end{figure}%
	\begin{figure}[bth]%
		\begin{minipage}{0.5\textwidth}%
			\begin{overpic}[width=\linewidth]{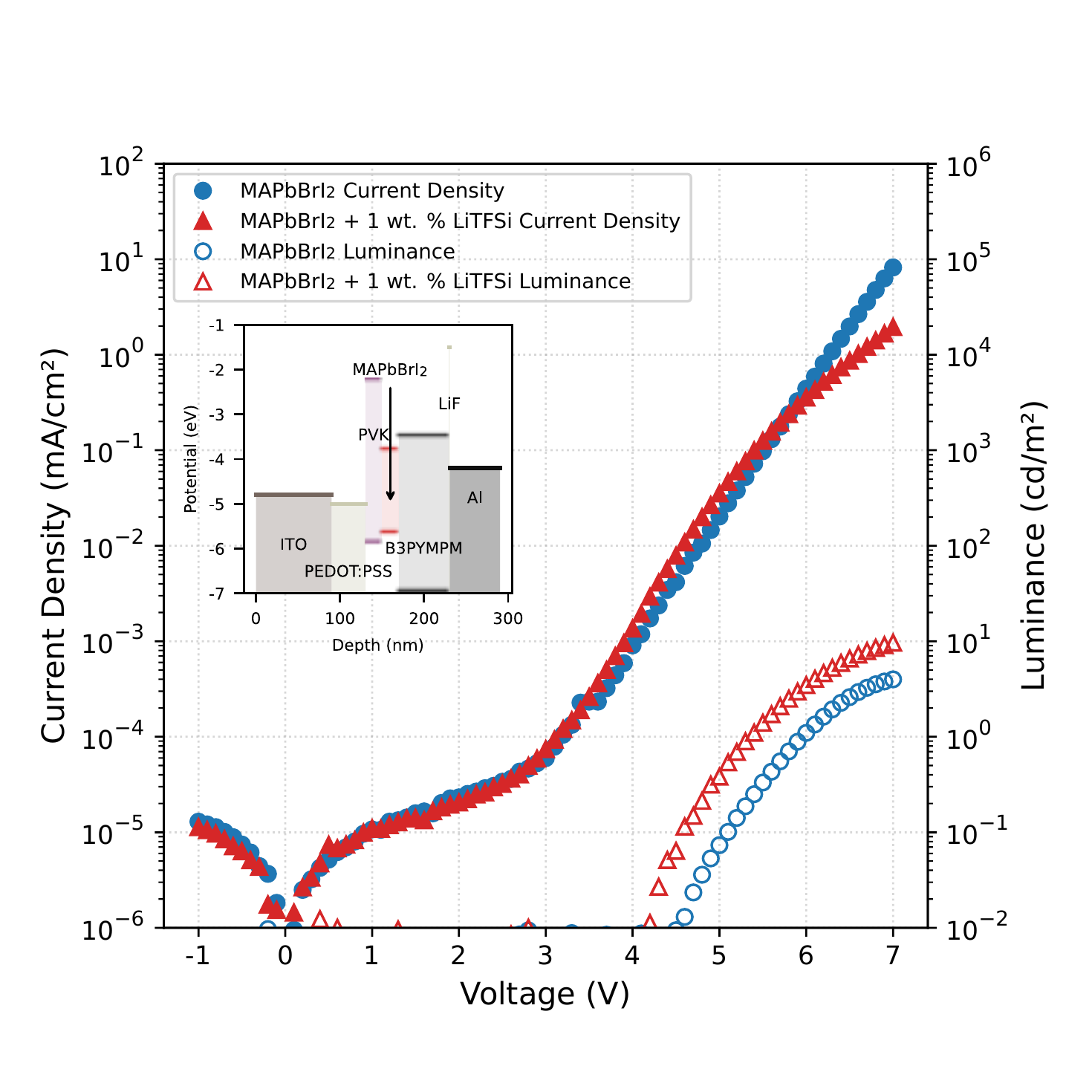}%
				\put (5,95) {\scriptsize a)}%
			\end{overpic}%
		\end{minipage}{}%
		\begin{minipage}{0.5\textwidth}%
			\begin{overpic}[width=\linewidth]{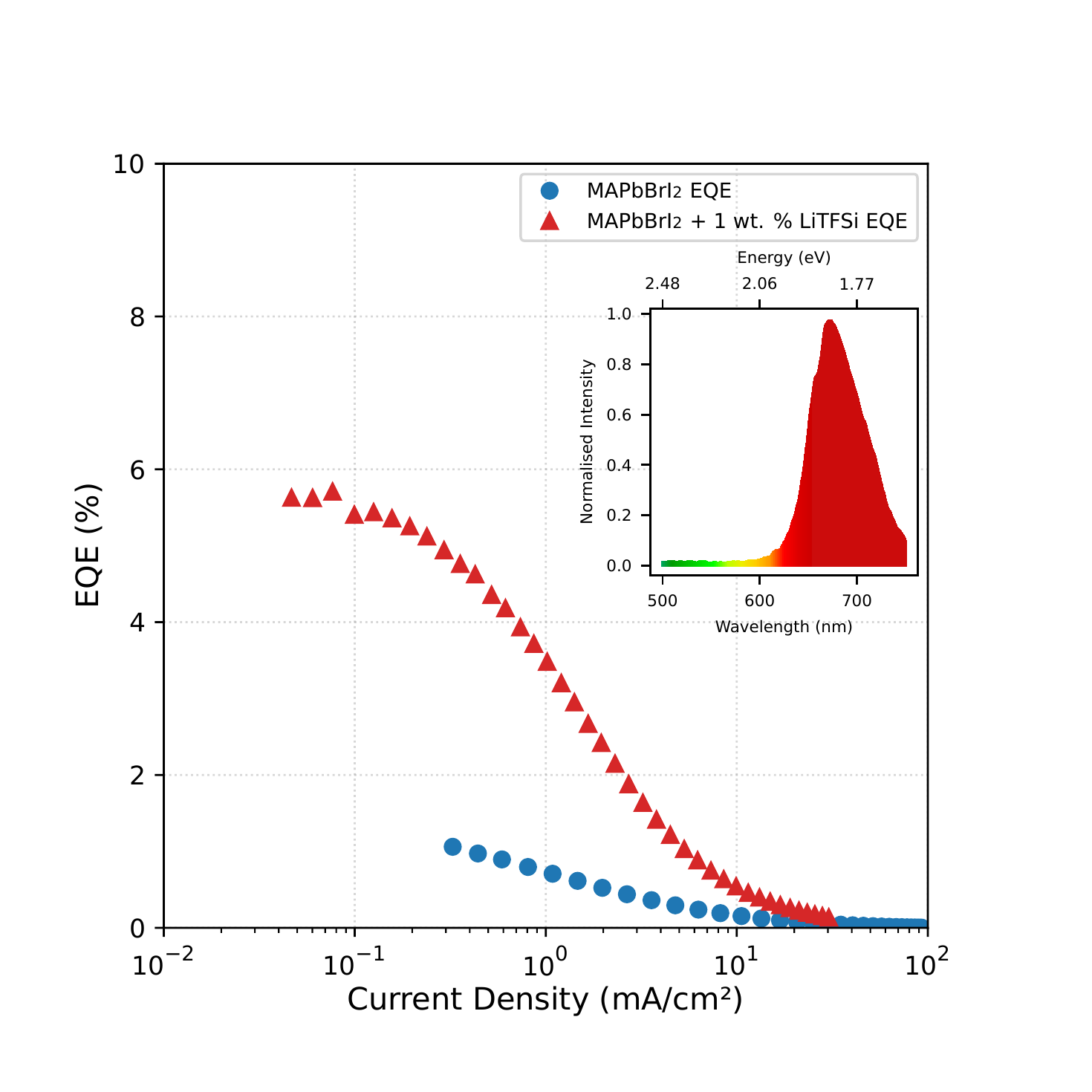}%
				\put (5,95) {\scriptsize b)}%
			\end{overpic}%
		\end{minipage}%
		\caption{%
			a): j-V-L curves of MAPbBrI\ts2-NC-LEDs with and without LiTFSI doping, with b) its corresponding EQE vs current density curves. %
			The insets depict their stack structures and their partly degraded emission spectra.%
		}%
		\label{rLED}%
	\end{figure}%
	\clearpage%
	\paragraph{Substrate cleaning procedure}{
	    All substrates are cleaned in a cleanroom by the same procedure: 4 sequential ultrasonic baths in different liquids for 5 minutes each is the initial cleaning step. %
	    The sequence of solvents is: Aloconox (detergent) enriched rinsing water, de-ionised water followed by UV-grade acetone and UV-grade isopropanol. %
	    After drying completely, the substrates are put in an UV-Ozone cleaner, for 15 minutes.%
	}%
	\paragraph{Additional Materials}%
	{
	    HATCN (Dipyrazino[2,3-f:2',3'-h] quinoxaline-2,3,6,7,10,11- hexacarbonitrile, sublimed\,>\,99\,\%, ProductID: LT-N221), B3PYMPM (4,6-Bis(3,5-di(pyridin-3-yl)phenyl)-2-methylpyrimidine, sublimed\,>\,99\,\%, ProductID: LT-N876) and LiF (Lithium Fluoride, >\,99,99\,\%, ProductID: LT-E001) have been obtained from Luminescence Technology Corp. (Lumtec, Taiwan). %
	    1-Octadecene (ODE), technical grade, 90\,\%, Sigma Aldrich; Oleic acid (OA), 97\,\%, Acros Organics; Oleylamine (OAm), 80\,-\,90\,\%, Acros Organics; Caesium carbonate (Cs\ts2CO\ts3), 99.99\,\% (trace metal basis), Acros Organics; Lead(II) oxide (PbO), 99.999\,\% (trace metal basis), Sigma Aldrich; Toluene, 99.8\,\%, extra dry, AcroSeal, Acros; Acetonitrile (ACN) , 99.9\,\%, extra dry, AcroSeal, Acros; Lead(II) iodide (PbI\ts2), 99\,\%, Acros Organics; Methylammonium bromide (MABr), 98\,\%, Sigma Aldrich; Methylamine (MA) solution, 33\,wt.\,\% in absolute Ethanol, Sigma Aldrich; Disodium ethylenediaminetetraacetic acid dihydrate (Na\ts2EDTA), Sigma Aldrich; All chemicals were used as purchased. 
	}%
	\paragraph{MAPbBrI\ts{2} nanocystals preparation}{%
	    The perovskite precursor solution has been prepared exactly as mentioned by Hassan et. al..\cite{Hassan2019} %
	    In order to obtain 12\,nm CH\ts3NH\ts3PbBrI\ts2 perovskite nanoparticles emitting at 650\,nm, 5\,ml of anhydrous toluene has been mixed with 4\,ml of oleic acid and 0.4\,ml of oleylamine in a three-neck flask under nitrogen atmosphere at 70\,\textdegree C. %
	    At this temperature 0.4\,ml of the previously made ACN/MA perovskite precursor solution has been swiftly injected into the toluene/ligand mixture under vigorous stirring. %
	    After one minute, the reaction vessel has been cooled to room temperature using an ice-bath. %
	    MAPbBrI\ts2 NCs have been collected by centrifuging the suspension without addition of an antisolvent (7000\,rpm, 10\,min.), decanting the supernatant, and collecting the precipitate. %
	    The precipitate has been centrifuged again without addition of a solvent (7000\,rpm, 5\,min), and the resulting supernatant has been removed with a syringe, to separate the traces of residual supernatant. %
	    The precipitate has been dissolved in 5\,ml anhydrous toluene and centrifuged again (7000\,rpm, 5\,min) to remove aggregates and larger particles. %
	    The resulting supernatant has been filtered through a 0.2\,\textmu m PTFE syringe filter and stored as stock solution inside of a glovebox. %
	    The as obtained MAPbBrI\ts2 solutions have been subsequently treated with Na\ts2ETDA. %
	    The solid EDTA salt has been added to the toluene solution of NCs and stirred for 24 hours. %
	    In a typical exchange reaction 10 times the amount of salt was used.\cite{Hassan2021} %
	    Afterwards, the samples have been filtered with a 0.2\,\textmu m syringe filter.%
	}%
	\clearpage%
	\bibliography{2022_Naujoks_QE_Enhancement_of_LHP_LEDs_by_LiTFSI_acsami.bib}%